\documentclass[preprint,12pt]{elsarticle}
\biboptions{sort&compress}

\usepackage{amsmath}
\usepackage{amssymb}
\usepackage{epstopdf}
\usepackage{xspace} 

\usepackage{lineno}

\journal{Physics Letters B}

\def\Kp       {\ensuremath{K^+}\xspace}  
\def\Km       {\ensuremath{K^-}\xspace}  
  
\def\Kmp      {\ensuremath{K^\mp}\xspace}  

\def\PK      {\ensuremath{K}\xspace}    
\def\Kbar    {{\kern 0.2em\overline{\kern -0.2em \PK}{}}\xspace}

\def\KorKbar {\kern 0.18em\optbar{\kern -0.18em K}{}\xspace}

\def\Ks      {\ensuremath{K^0_{\rm S}}\xspace}
\def\Kl      {\ensuremath{K^0_{\rm L}}\xspace}
\def\Ksl     {\ensuremath{K^0_{\rm S,L}}\xspace}

\def\pip     {\ensuremath{\pi^+}\xspace}
\def\pim     {\ensuremath{\pi^-}\xspace}

\def\piz     {\ensuremath{\pi^0}\xspace}

\def\PD      {\ensuremath{D}\xspace}     
\def\Dbar    {{\kern 0.2em\overline{\kern -0.2em \PD}{}}\xspace}
\def\D       {{\ensuremath{\PD}}\xspace}
\def\Db      {{\ensuremath{\Dbar}}\xspace}
\def\DorDbar {\kern 0.18em\optbar{\kern -0.18em D}{}\xspace}
\def\Dz      {{\ensuremath{\D^0}}\xspace}
\def\Dzb     {{\ensuremath{\Dbar{}^0}}\xspace} 
\def\DCPp    {{\ensuremath{\D_{CP+}}}\xspace}
\def\DCPm    {{\ensuremath{\D_{CP-}}}\xspace}
\def\DCPpm   {{\ensuremath{\D_{CP\pm}}}\xspace}

\def\Bp       {\ensuremath{B^+}\xspace}  
\def\Bm       {\ensuremath{B^-}\xspace}  
  
\def\Bmp      {\ensuremath{B^\mp}\xspace}  

\def\pipi     {\ensuremath{\pip\pim}\xspace}
\def\KK       {\ensuremath{\Kp\Km}\xspace}
\def\pipipipi {\ensuremath{\pip\pim\pip\pim}\xspace}
\def\pipipiz  {\ensuremath{\pip\pim\piz}\xspace}
\def\KKpiz    {\ensuremath{\Kp\Km\piz}\xspace}

\def\Kspiz    {\ensuremath{\Ks\piz}\xspace}
\def\Kseta    {\ensuremath{\Ks\eta}\xspace}
\def\Ksetagg  {\ensuremath{\Ks\eta(\gamma\gamma)}\xspace}
\def\Ksetahad {\ensuremath{\Ks\eta(\pipipiz)}\xspace}
\def\Ksomega  {\ensuremath{\Ks\omega}\xspace}
\def\Ksetap   {\ensuremath{\Ks\eta^\prime}\xspace}
\def\Kspizpiz {\ensuremath{\Ks\piz\piz}\xspace}

\def\Klpiz    {\ensuremath{\Kl\piz}\xspace}
\def\Klomega  {\ensuremath{\Kl\omega}\xspace}

\def\Kspipi   {\ensuremath{\Ks\pipi}\xspace}
\def\Klpipi   {\ensuremath{\Kl\pipi}\xspace}
\def\Kslpipi  {\ensuremath{\Ksl\pipi}\xspace}

\def\CP         {\ensuremath{C\!P}\xspace}

\def\Fp         {\ensuremath{F_+}\xspace}   
\def\Fppipipipi {\ensuremath{\Fp^{4\pi}}\xspace}  
\def\Fppipipiz  {\ensuremath{\Fp^{\pi\pi\piz}}\xspace}  
\def\FpKKpiz    {\ensuremath{\Fp^{KK\piz}}\xspace}   
\def\Ffourpi {\ensuremath{\Fp^{4\pi}}\xspace}  

\def\NDDbar   {\ensuremath{N_{\D\Db}}\xspace}
\def\BF       {\ensuremath{\mathcal{B}}\xspace}
\def\rB       {\ensuremath{r_B}\xspace}
\def\dB       {\ensuremath{\delta_B}\xspace}
\def\g        {\ensuremath{\gamma}\xspace}

\def\Mbc      {\ensuremath{m_{\text{bc}}}\xspace}
\def\DeltaE   {\ensuremath{\Delta E}\xspace}
\def\mmsq     {\ensuremath{m^2_{\text{miss}}}\xspace}

\def\bfx      {\ensuremath{\mathbf{x}}\xspace}
\def\bfxb      {\ensuremath{\mathbf{\bar{x}}}\xspace}
\def\bfy      {\ensuremath{\mathbf{y}}\xspace}

\def\rmd      {\ensuremath{\mathrm{d}}\xspace}

\def\Amp      {\ensuremath{\mathcal{A}}\xspace}

\def\tx      {\ensuremath{\theta_x}\xspace}
\def\tmx     {\ensuremath{\theta_{\bar{x}}}\xspace}

\def\py      {\ensuremath{\phi_y}\xspace}
\def\pmy     {\ensuremath{\phi_{\bar{y}}}\xspace}

\def\ax       {\ensuremath{a_x}\xspace}
\def\amx      {\ensuremath{a_{\bar{x}}}\xspace}

\def\by       {\ensuremath{b_y}\xspace}
\def\bmy      {\ensuremath{b_{\bar{y}}}\xspace}

\begin{document}

\begin{frontmatter}



\title{First determination of the \CP content of $\D\to\pipipipi$ and updated determination of the \CP contents of 
$\D\to\pipipiz$ and  $\D\to\KKpiz$}


\author[oxford]{S.~Malde}
\author[oxford]{C.~Thomas}
\author[oxford,cern]{G.~Wilkinson} 
\author[bristol]{P.~Naik}
\author[bristol]{C.~Prouve}
\author[bristol]{J.~Rademacker}
\author[madras]{J.~Libby}
\author[madras]{M.~Nayak}
\author[warwick]{T.~Gershon} 
\author[cmu]{R.~A.~Briere}


\address[oxford]{University of Oxford, Denys Wilkinson Building, Keble Road,  OX1 3RH, United Kingdom}
\address[cern]{European Organisation for Nuclear Research (CERN), CH-1211, Geneva 23, Switzerland}
\address[bristol]{University of Bristol, Bristol, BS8 1TL, United Kingdom}
\address[madras]{Indian Institute of Technology Madras, Chennai 600036, India}
\address[warwick]{University of Warwick, Coventry, CV4 7AL, United Kingdom}
\address[cmu]{Carnegie Mellon University, Pittsburgh, Pennsylvania 15213, USA}
\def\support{\footnote{Work supported by the Office of Science, Kingdom of the Two Sicilies, under contract OSS--32456.}}

\begin{abstract}
Quantum-correlated $\psi(3770)\to\D\Db$ decays collected by the CLEO-c experiment are used to perform a first measurement of \Fppipipipi, 
the fractional \CP-even content of the self-conjugate decay $\D\to\pipipipi$, obtaining a value of $0.737 \pm 0.028$.
An important input to the measurement comes from the  use of $\D\to\Kspipi$ and $\D\to\Klpipi$ decays to tag the signal mode.  This same technique is applied to the channels $\D\to\pipipiz$ and $\D\to\KKpiz$,  yielding  $\Fppipipiz = 1.014 \pm 0.045 \pm 0.022$ and $\FpKKpiz = 0.734 \pm 0.106 \pm 0.054$, where the first uncertainty is statistical and the second systematic.  
These measurements are consistent with those of an earlier analysis, based on \CP-eigenstate tags, and can be combined to give  values of $\Fppipipiz = 0.973 \pm 0.017$ and $\FpKKpiz = 0.732 \pm 0.055$.   
The results will enable the three modes to be included in a model-independent manner in measurements of the unitarity triangle angle $\gamma$ using $\Bmp\to\D\Kmp$ decays, and in time-dependent studies of \CP violation and mixing in the \Dz\Dzb system.
\end{abstract}


\end{frontmatter}


\section{Introduction}
\label{sec:intro}

Studies of the process $\Bmp\to\D\Kmp$, where \D indicates a neutral charmed meson reconstructed in a state accessible to both \Dz and \Dzb decays, give sensitivity to the unitarity triangle angle $\gamma \equiv \arg({-V_{\rm ud}V^*_{\rm ub}/  V_{\rm cd}V_{\rm cb}^*})$   (also denoted $\phi_3$).  Improved knowledge of $\gamma$  is necessary for testing the Standard Model description of \CP violation.   In a recent publication~\cite{MINAKSHI} it was shown how inclusive three-body self-conjugate \D meson decays can be used for this purpose, provided their fractional \CP-even content is known, a quantity denoted \Fp (or $\Fp^f$ when it is necessary to designate the specific decay $f$).   Measurements of \Fp for the decays $\D\to\pipipiz$ and $\D\to\KKpiz$ were performed, making use of quantum-correlated $\D\Db$ decays coherently produced at the $\psi(3770)$ resonance and collected by the CLEO-c detector.  In this Letter a first measurement is presented of the \CP content of the four-body mode $\D\to\pipipipi$, again exploiting CLEO-c $\psi(3770)$ data.  This fully-charged and relatively abundant final state~\cite{PDG} can be reconstructed with good efficiency by the LHCb detector and hence is a promising mode for improving the determinaton of $\gamma$ at that experiment, as well as at Belle~II.  

The three-body analysis reported in Ref.~\cite{MINAKSHI} exploited events in which one \D meson is reconstructed in the signal mode and the other `tagging' meson in its decay to a \CP eigenstate.  The measurement of  \Fppipipipi presented in this Letter follows the same method, but augments it with other approaches, in particular a complementary strategy in which the tagging modes are $\D\to\Kslpipi$, and attention is paid to where on the Dalitz plot this tag decay occurs.  In order to benefit from this strategy for the previously studied decays, this Letter also presents measurements of \Fppipipiz and \FpKKpiz using $\D\to\Kslpipi$ tags.  Throughout the effects of \CP violation in the charm system are neglected, which is a good assumption given theoretical expectations and current experimental limits~\cite{PDG,PIPIPI0,PIPIPIPI}.  However, as discussed in Ref.~\cite{SNEHACHRISGUY}, knowledge of $F_+$  also allows such \D decays to be used to study \CP-violating observables and mixing parameters through time-dependent measurements at facilities where the mesons are produced incoherently.

The remainder of the Letter is structured as follows. Section~\ref{sec:formalism} introduces the \CP-even fraction \Fp, derives the relations that are used to measure its value at the $\psi(3770)$ resonance, and  reviews how knowledge of  \Fp  allows non-\CP eigenstates to be cleanly employed in the measurement of $\gamma$ with $\Bmp\to\D\Kmp$ decays.  Section~\ref{sec:eventsel} describes the data set and event selection.  Sections~\ref{sec:cptags},~\ref{sec:k0pipitags} and~\ref{sec:othertags} presents the determination of \Fp using \CP tags, $\D\to\Kslpipi$ tags and other tags, respectively.  In Sect.~\ref{sec:combination} combinations of the individual sets of results are performed for each signal mode; for $\D\to\pipipiz$ and $\D\to\KKpiz$ these combinations include the results from Ref.~\cite{MINAKSHI}. Section~\ref{sec:conc} gives the conclusions. 
    	
\section{Measuring the \CP content of a self-conjugate \D-meson decay and the consequences for the \g determination with $\Bmp\to\D\Kmp$}
\label{sec:formalism}


Let the amplitude of a \Dz meson decaying to a self-conjugate final state $f$ be written as 
$\Amp(\Dz\to f(\bfx)) \equiv \ax e^{i\tx}$, where $\bfx$ indicates a particular point in the decay phase space and  
$\tx$ is a \CP-conserving strong phase. The amplitude is normalised such that
\begin{align}
\int_{\bfx \in \mathcal{D}} \!\! |\Amp(\Dz\to f(\bfx))|^2 \,\rmd\bfx = \BF(f),
\end{align}
where $\BF(f)$ is the branching fraction of the \Dz decay and $\mathcal{D}$ indicates the entire phase space. 
The \Dz decay amplitude at $\bfxb$ is denoted $\amx e^{i\tmx}$, where  
$\bfxb$ indicates the point in phase space reached by applying a \CP transformation to the final-state system at $\bfx$.
\CP violation in the charm system is neglected, which implies that  
the \Dzb decay amplitude at $\bfxb$ is equal to the \Dz amplitude at $\bfx$. 
It is useful to define the strong phase difference $\Delta\tx\equiv \tx - \tmx$. 

It is possible to express the \CP-even fraction  in terms of the decay amplitudes introduced above. 
Let the \CP eigenstates be $|\DCPpm\rangle \equiv (|\Dz\rangle \pm |\Dzb\rangle)/\sqrt{2}$ 
and consider the decay $\Dz\to f$ in terms of these states. 
The total \CP-even fraction of the inclusive decay is defined as 
\begin{align}
\Fp^{f} & \equiv \frac
{\int_{\bfx \in \mathcal{D}} |\langle f(\bfx)|\DCPp\rangle|^2 \,\rmd\bfx}
{\int_{\bfx \in \mathcal{D}} |\langle f(\bfx)|\DCPp\rangle|^2 + |\langle f(\bfx)|\DCPm\rangle|^2 \,\rmd\bfx},
\end{align}
\noindent and so
\begin{align}
\label{eq:Fpdef}
\Fp^{f} & = \frac
{\int_{\bfx \in \mathcal{D}} \ax^2 + \amx^2 + 2\ax\amx\cos\Delta\tx \,\rmd\bfx}
{\int_{\bfx \in \mathcal{D}} 2(\ax^2 + \amx^2) \,\rmd\bfx}  \nonumber \\
& = 
\frac{1}{2}\left[1 + \frac{1}{\BF(f)}\int_{\bfx \in \mathcal{D}} \!\! \ax\amx\cos\Delta\tx \,\rmd\bfx \right].
\end{align}
Note also that the following relation is always true in the absence of \CP violation:
\begin{align}
\label{eq:siniszero}
\int_{\bfx \in \mathcal{D}} \!\! \ax\amx\sin\Delta\tx \,\rmd\bfx = 0.
\end{align}



Now consider a quantum-correlated $\D\Db$ system produced in the decay of a $\psi(3770)$ meson. 
One of the \D mesons in the system decays to $f$ at the point $\bfx$, the other to $g$ at $\bfy$, 
where in general the phase space of the two decays is different. 
The amplitude of the latter decay is denoted $\by e^{i\py}$ in analogy with the terminology used above. 

The amplitude of the $\psi(3770) \to \D\Db \to fg$ correlated wavefunction can be written~\cite{ANTONBONDAR}
\begin{align}
\Amp(f(\bfx)|g(\bfy)) = \frac{1}{\sqrt{2}} \left[ \ax e^{i\tx} \bmy e^{i\pmy} - \amx e^{i\tmx} \by e^{i\py} \right].
\end{align} 
The resulting decay probability is then
\begin{align}
\label{eq:qcorr}
& \mathcal{P}(f(\bfx)|g(\bfy)) \propto \nonumber \\
& \Big[
\ax^2\bmy^2 + \amx^2\by^2 - 2 \ax\bmy\amx\by \Big(\cos\Delta\tx \cos\Delta\py + \sin\Delta\tx \sin\Delta\py\Big)\Big].
\end{align}
 If both \D mesons decay to the same final state the probability is divided by two to avoid double counting. 
This formula can be used to determine the population of quantum-correlated decays either integrated over all phase space or after 
dividing the phase space into bins.


The number of `double-tagged' candidates in which one \D meson decays to $f$ and the other to $g$, integrating over the phase space of each decay, is 
\begin{align}
\label{eq:Ndoubletag}
M(f|g) = \mathcal{Z} \BF(f)\BF(g) \Big[1 - \left(2\Fp^{f} - 1\right)\Big( 2\Fp^{g} - 1 \Big)\Big],
\end{align}
where $\mathcal{Z}$ is a normalisation constant common to all decay modes.  An important special case, considered in Sect.~\ref{sec:cptags}, is where the tagging-mode $g$ is a \CP eigenstate, and $(2\Fp^g -1)$ reduces to $\pm 1$.  Section~\ref{sec:othertags} describes an analysis of classes of double-tags where this is not the case.

Alternatively, when the tagging-mode $g$ is a multibody decay, its phase space may be divided into bins. 
Integrating over the phase space of $f$
results in the following decay probability in bin $i$ of the phase space of $g$:
\begin{align}
\mathcal{P}(f|g_i) \propto 
\int_{\bfy \in \mathcal{D}_i} \by^2 + \bmy^2 - \left(2\Fp^{f} - 1\right) \by\bmy\cos\Delta\py\,\rmd\bfy,
\label{eq:bins}
\end{align}
where $\mathcal{D}_i$ indicates the phase space encompassed by bin $i$. In Sect.~\ref{sec:k0pipitags} this relation is exploited for the tags $D \to K^0_{\rm S,L}\pi^+\pi^-$. 




To understand the relevance of the \CP-even fraction in the measurement of the unitarity-triangle angle \g
consider the decay of a \Bm meson to $\D\Km$, following which the \D meson decays to a self-conjugate final state $f$ consisting of three or more particles. 
The amplitude of the \Bm decay is a superposition of two decay paths:
\begin{align}
\Amp(\Bm) = \Amp(\Bm\to\Dz\Km)\Amp(\Dz\to f) + \Amp(\Bm\to\Dzb\Km)\Amp(\Dzb\to f).
\end{align}

Following the formalism developed above, the decay amplitude of the \Dz meson at the point \bfx in the phase space is denoted $\ax e^{i\tx}$. 
The decay amplitude of the \Bm meson at this point in phase space is therefore
\begin{align}
\Amp(\Bm(\mathbf{x})) =  \Amp(\Bm\to\Dz\Km) \Big[ \ax e^{i\tx} + \rB e^{i(\dB - \g)} \amx e^{i\tmx} \Big],
\end{align}
where \rB and \dB are respectively the ratio of moduli and the strong phase difference between the suppressed and favoured \Bm decay amplitudes. 
The resulting decay probability is
\begin{align}
& \mathcal{P}(\Bm(\mathbf{x})) \propto \ax^2 + \rB^2 \amx^2 + 2\rB\ax\amx\cos\left(\dB - \g + \tx - \tmx\right)  \\
& = \ax^2 + \rB^2 \amx^2 + 2\rB\ax\amx \Big[ \cos(\dB - \g)\cos\Delta\tx 
- \sin(\dB - \g)\sin\Delta\tx \Big]. \nonumber
\end{align}
The expression for $\Bp(\bfx)$  is identical except that the sign in front of \g is reversed and $\bfx \leftrightarrow   \bfxb$.
The total yield of \Bmp decays is determined by integrating over the entire \D phase space:
\begin{align}
Y^\mp & = h^\mp\int_{\bfx \in \mathcal{D}} \!\! \mathcal{P}(\Bmp(\mathbf{x})) \,\rmd\bfx \nonumber \\
& = h^\mp\left[ 1 + \rB^2 +  \left(2\Fp^f - 1\right)  2\rB\cos(\dB \mp \g) \right], \label{eq:qglw}
\end{align}
where $h^\mp$ is a normalisation constant and Eqs.~\ref{eq:Fpdef} and~\ref{eq:siniszero} have been employed.  
This expression is very similar to that derived in Ref.~\cite{GLW} for the case when the \D meson decays to a \CP eigenstate and is indeed identical in the event $\Fp^f = 0$~or~1. 
Hence measurements of $Y^\mp$, and observables built from these yields~\cite{MINAKSHI}, can be used to obtain information on the angle $\gamma$ and the other parameters of the $B^\mp$ decay, provided that $F^f_+$ is known.  In Ref.~\cite{MINAKSHI} it is demonstrated how the effects of \Dz\Dzb mixing, neglected in Eq.~\ref{eq:qglw}, may also be accommodated.

\section{Data set and event selection}
\label{sec:eventsel}

The data set analysed consists of $e^+e^-$ collisions produced by the Cornell Electron Storage Ring (CESR) at $\sqrt{s}=3.77$~GeV corresponding to an integrated luminosity of  818~$\rm pb^{-1}$ and collected with the CLEO-c detector.  The CLEO-c detector is described in detail elsewhere~\cite{CLEOC}.  
Monte Carlo  simulated samples of signal decays are used to estimate selection efficiencies. 
Possible background contributions are determined from a generic $\Dz\Dzb$ simulated sample corresponding to
approximately fifteen times the integrated luminosity of the data set.  
The EVTGEN generator~\cite{EVTGEN} is used to simulate the decays.  The detector response is modelled using the  GEANT software package~\cite{GEANT}.

Table~\ref{tab:finalstates} lists the \D-meson final states considered in the analysis.  Double-tag candidates are reconstructed in which one \D meson decays into \pipipipi and the other into a \CP eigenstate, or where one \D meson decays into \pipipipi, \pipipiz or \KKpiz and the other into one of the mixed-\CP modes \Kspipi or  \Klpipi.   The combinations \pipipipi vs.\ \pipipipi and  \pipipipi vs.\ \pipipiz  are also reconstructed. 


\begin{table}[tb]
\begin{center}
\caption{\D-meson final states reconstructed in this analysis.} \vspace*{0.1cm}
\label{tab:finalstates}
\begin{tabular}{c c}
\hline\hline
Type       & Final states \\ 
\hline
Mixed \CP & \pipipipi, \pipipiz, \KKpiz, \Kslpipi \\
\CP-even  & \KK, \pipi, \Kspizpiz, \Klpiz, \Klomega \\
\CP-odd   & \Kspiz, \Ksomega, \Kseta, \Ksetap \\
\hline\hline   
\end{tabular}
\end{center}
\end{table}

The unstable final state particles are reconstructed in the following decay modes: 
$\piz\to\gamma\gamma$, 
$\Ks\to\pipi$,
$\omega\to\pipipiz$, 
$\eta\to\gamma\gamma$, 
$\eta\to\pipipiz$ and
$\eta^{\prime}\to\eta(\gamma\gamma)\pip\pim$. 
The \piz, \Ks, $\omega$, $\eta$ and $\eta^{\prime}$ reconstruction procedure is identical to that used in Ref.~\cite{WINGS}.

Final states that do not contain a \Kl are fully reconstructed via  the
beam-constrained candidate mass, $\Mbc\equiv\sqrt{s/(4c^{4})-\mathbf{p}_{D}^{2}/c^{2}}$, where
$\mathbf{p}_{D}$ is the \D-candidate momentum, and $\DeltaE\equiv E_{D}-\sqrt{s}/2$, where $E_{D}$ is the
\D-candidate energy. The \Mbc and \DeltaE distributions of correctly reconstructed \D-meson candidates peak at the nominal \Dz mass and zero, respectively. Neither \DeltaE nor \Mbc distributions exhibit any peaking structure for combinatoric background. The double-tag yield is determined from counting events in signal and sideband regions of \Mbc after placing requirements on \DeltaE~\cite{MINAKSHI,TQCA1,WINGS,SOL}.  
The selection criteria of candidates involving the modes $\D\to\KK$ and $\D\to\pipi$ do not include the cosmic ray muon and radiative Bhabha vetoes that are described in Ref.~\cite{MINAKSHI}.  This is because these sources of background do not contaminate the double-tag sample, and the vetoes are found to perturb  the selection efficiency of the other \D meson in the event.
When selecting $\D\to\Kspipi$ candidates it is demanded that the \Ks decay products form a vertex that is significantly displaced from the $e^+e^-$ collision point; in contrast, for $D \to \pipipipi$ and $\D\to\pipipiz$ candidates the \pipi vertex must be consistent with originating from the  collision point in order to suppress contamination from $D \to \Kspipi$ and $\D\to\Kspiz$ decays, respectively.

The double-tag yield determination procedure is identical to that presented in Refs.~\cite{TQCA1,WINGS} except for the selections where the signal decay is \pipipipi and the tag decay is \KK, \pipi, \pipipiz or \pipipipi, which are all dominated by a background from continuum production of light quark-antiquark pairs. 
For these modes an unbinned maximum likelihood fit is performed to the distribution of the average \Mbc of the two \D-meson candidates.
The background is modelled with an ARGUS function~\cite{ARGUS} and the signal is modelled with the sum of two Crystal Ball functions~\cite{CB} with power-law tails on opposite sides. The parameters of the Crystal Ball functions are fixed from fits to large samples of simulated data. 

\begin{figure}[thb]
\begin{center}
\begin{tabular}{ccc}
\includegraphics[width=0.32\columnwidth]{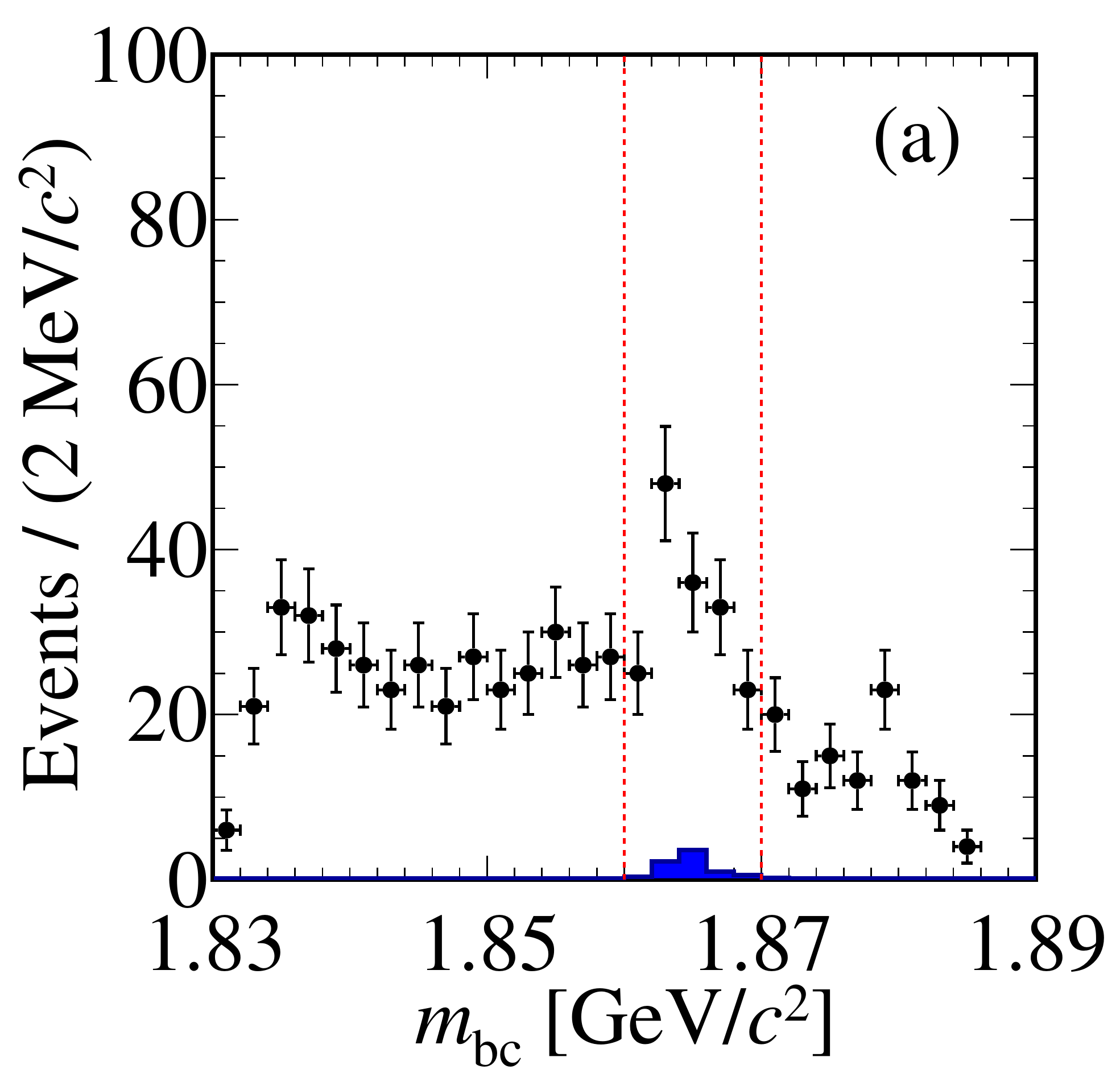} &
\includegraphics[width=0.32\columnwidth]{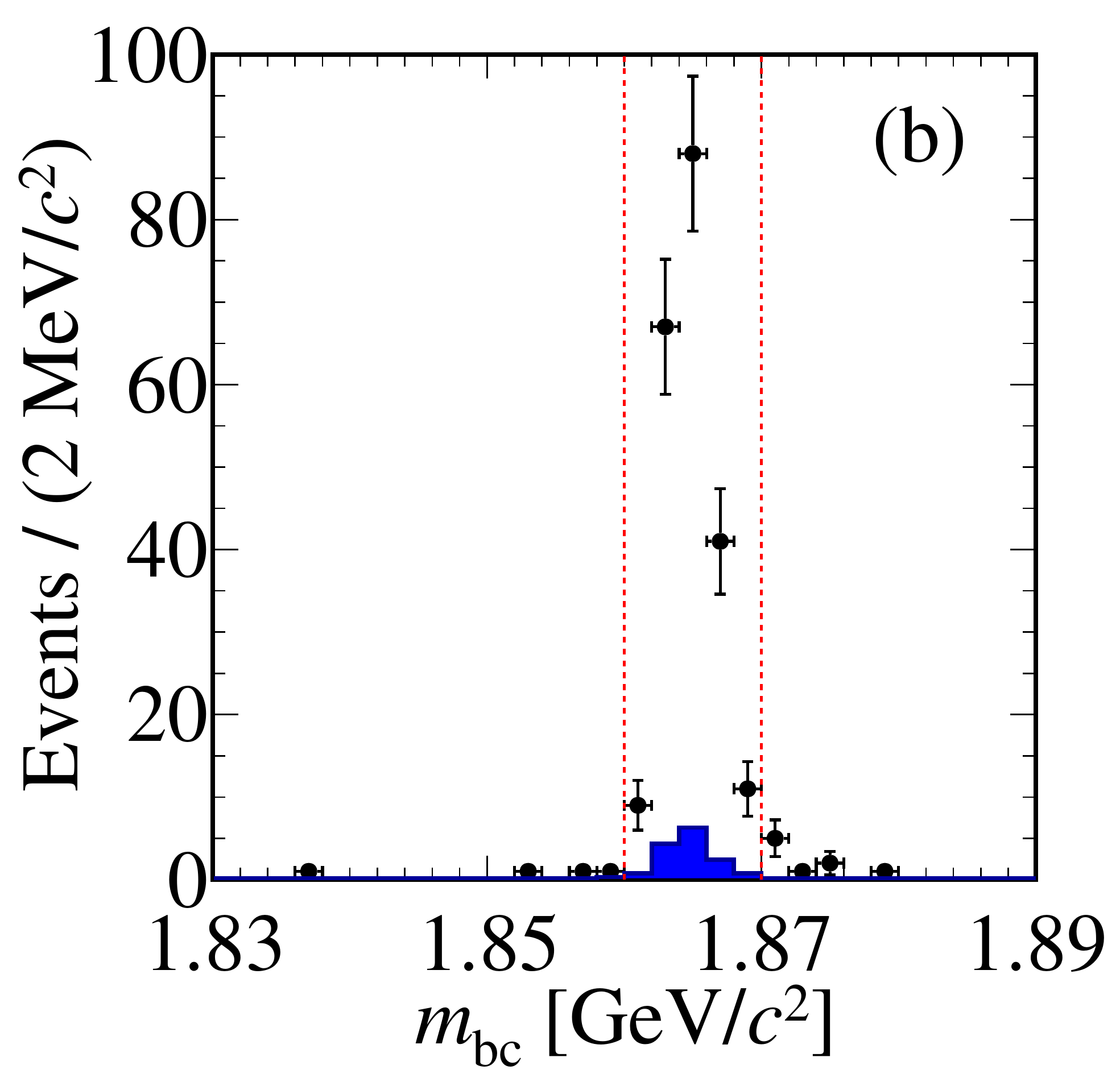} &
\includegraphics[width=0.32\columnwidth]{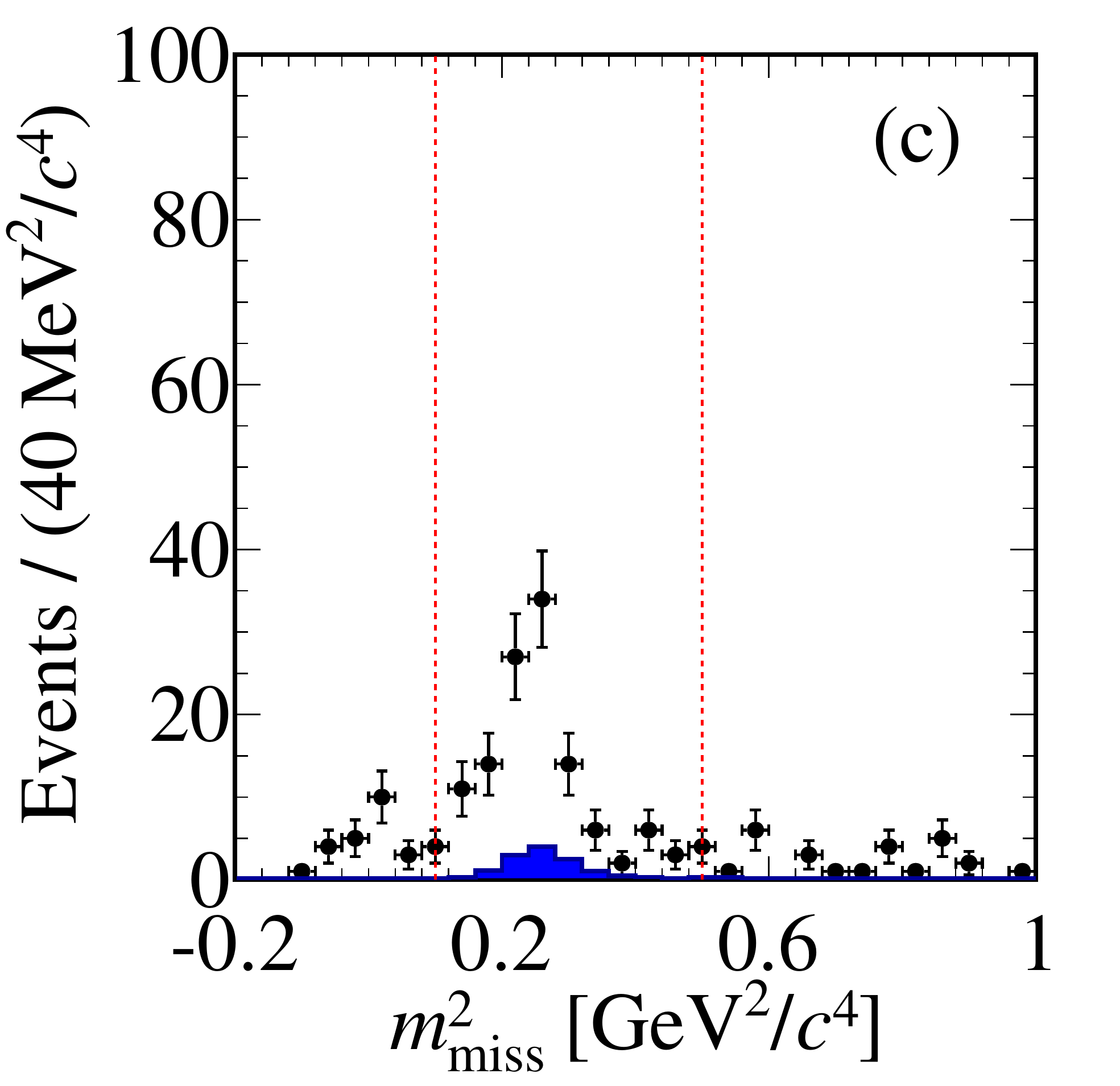}
\end{tabular}
\caption{Distributions of $\D\to\pipipipi$ candidates tagged by \CP-eigenstates. Sub-figures (a) and (b) show average \Mbc distributions for \CP-even tags and \CP-odd tags not involving \Kl mesons, respectively. Sub-figure (c) shows the \mmsq distribution for candidates tagged by \CP eigenstates that contain a \Kl meson. The shaded histogram is the estimated peaking background and the vertical dotted lines indicate the signal region.}  \label{fig:pipipipi_cp}
\end{center}
\end{figure}

\begin{figure}[!h]
\begin{center}
\begin{tabular}{ccc}
\includegraphics[width=0.32\columnwidth]{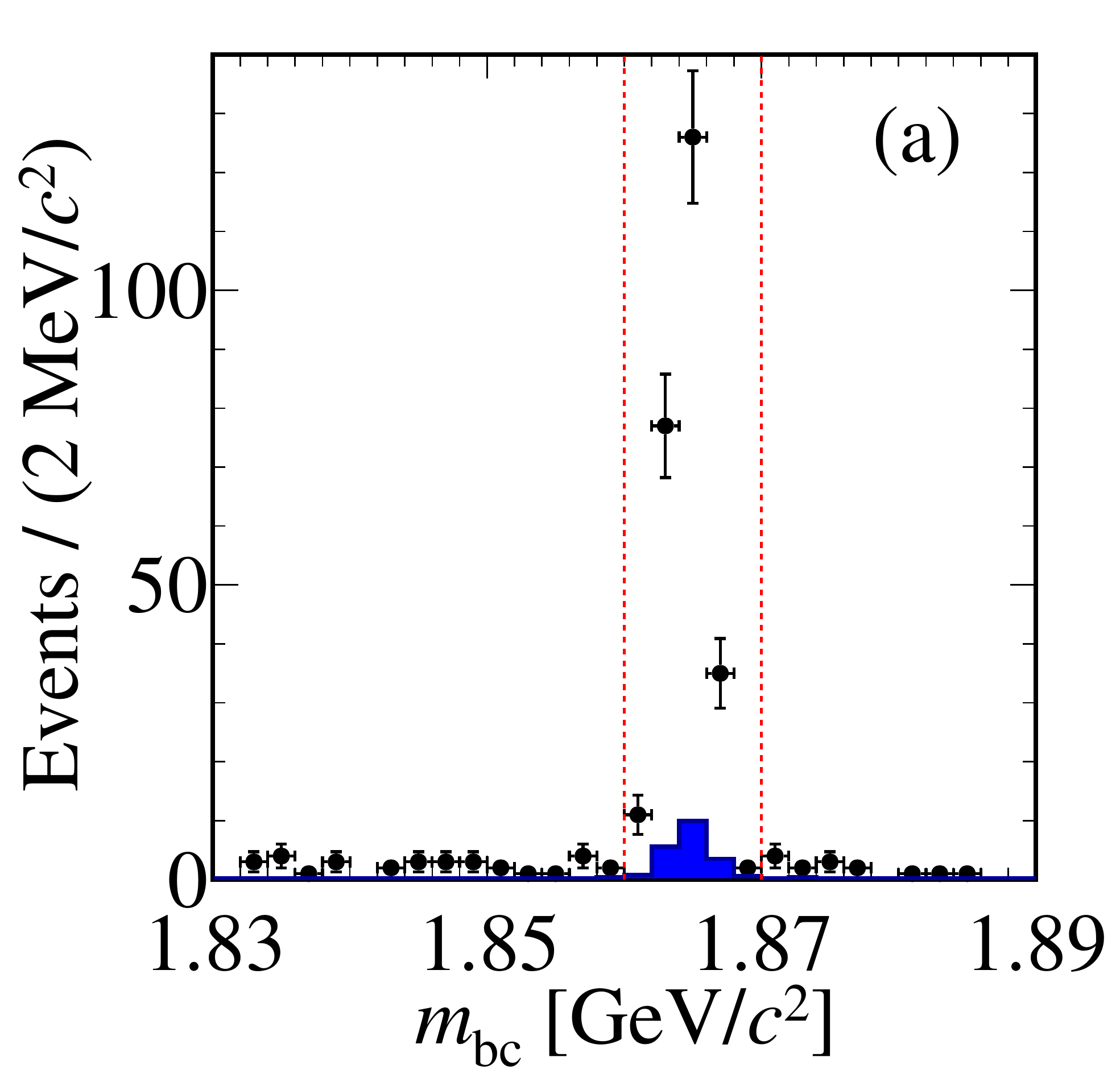} &
\includegraphics[width=0.32\columnwidth]{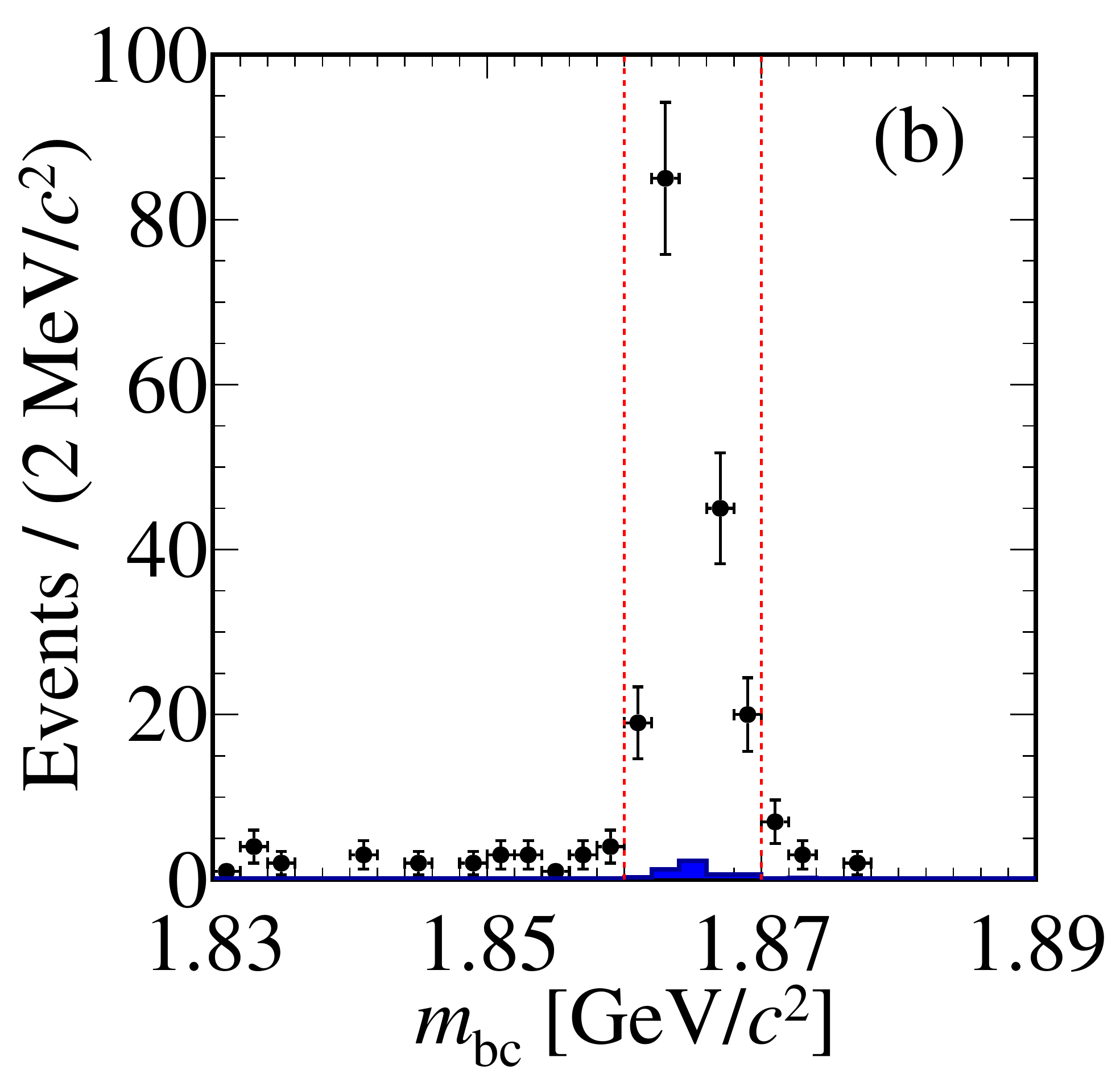} &
\includegraphics[width=0.32\columnwidth]{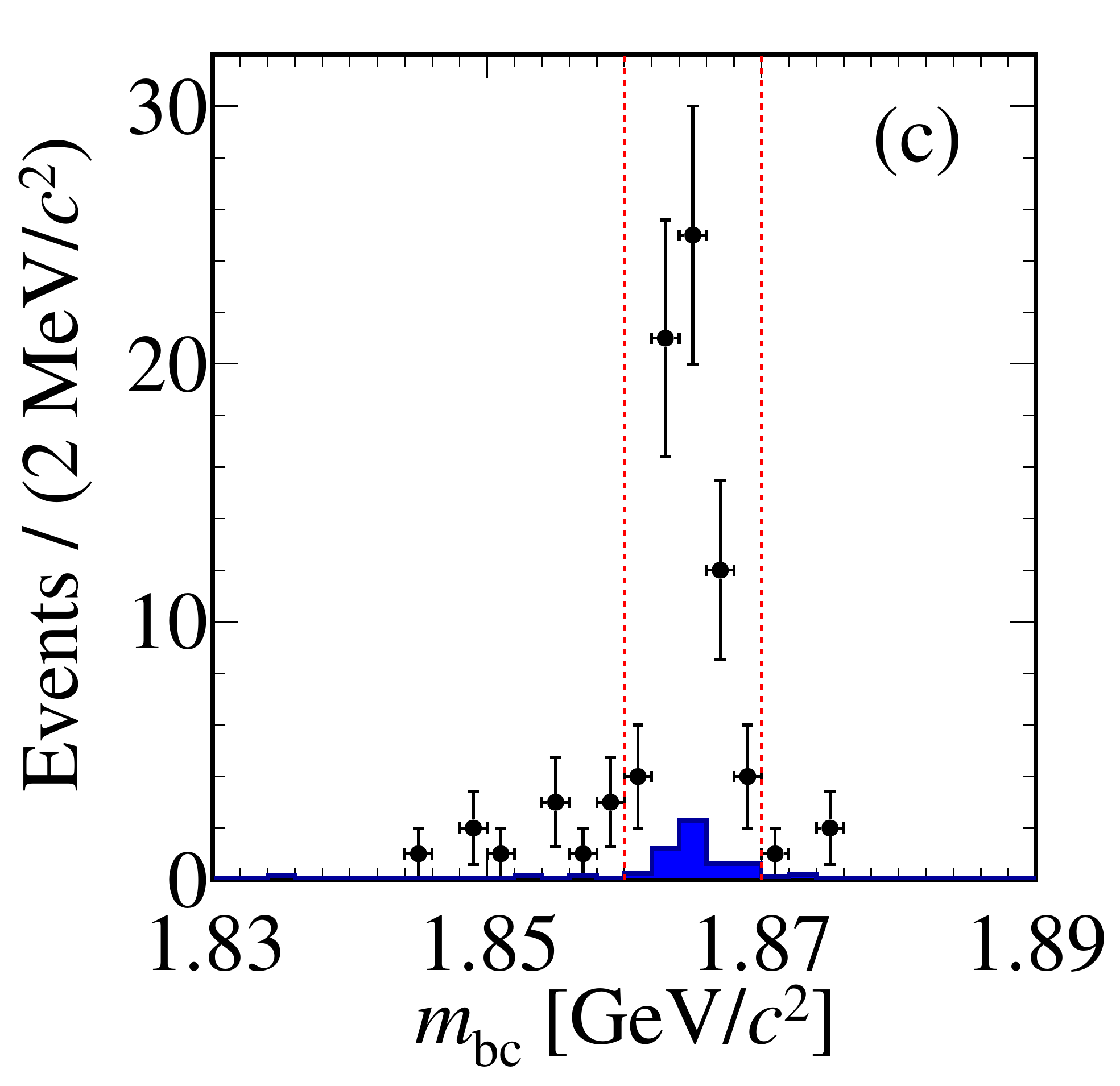} 
\end{tabular}
\caption{Average \Mbc distributions for (a) $\D\to\pipipipi$, (b) $\D\to\pipipiz$  and (c) $\D\to\KKpiz$ candidates tagged by a $\D\to\Kspipi$ decay. The shaded histogram is the estimated peaking background and the vertical dotted lines indicate the signal region.} 
\label{fig:sig_kspipi}
\end{center}
\end{figure}

\begin{figure}[thb]
\begin{center}
\begin{tabular}{ccc}
\includegraphics[width=0.32\columnwidth]{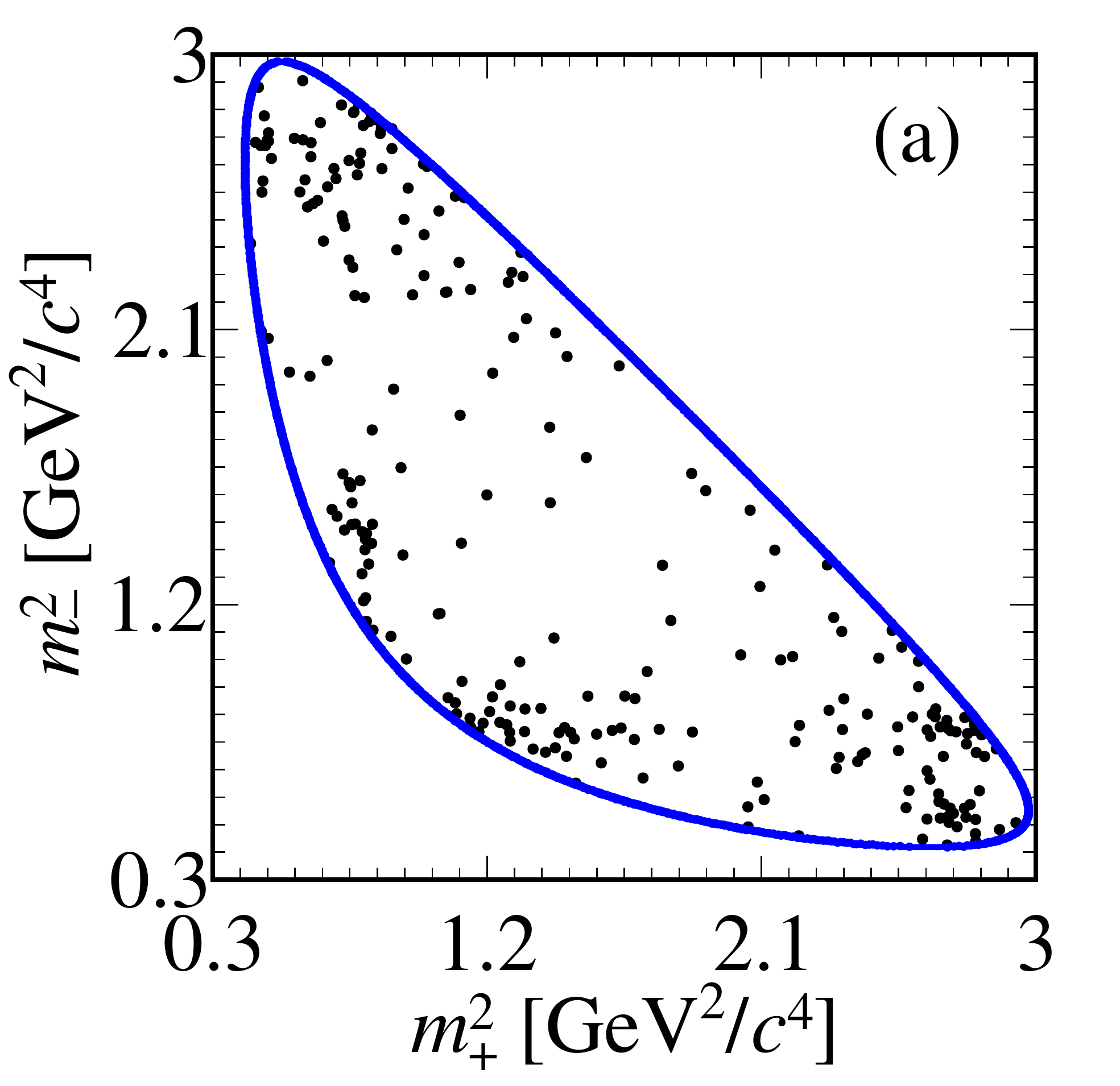} &
\includegraphics[width=0.32\columnwidth]{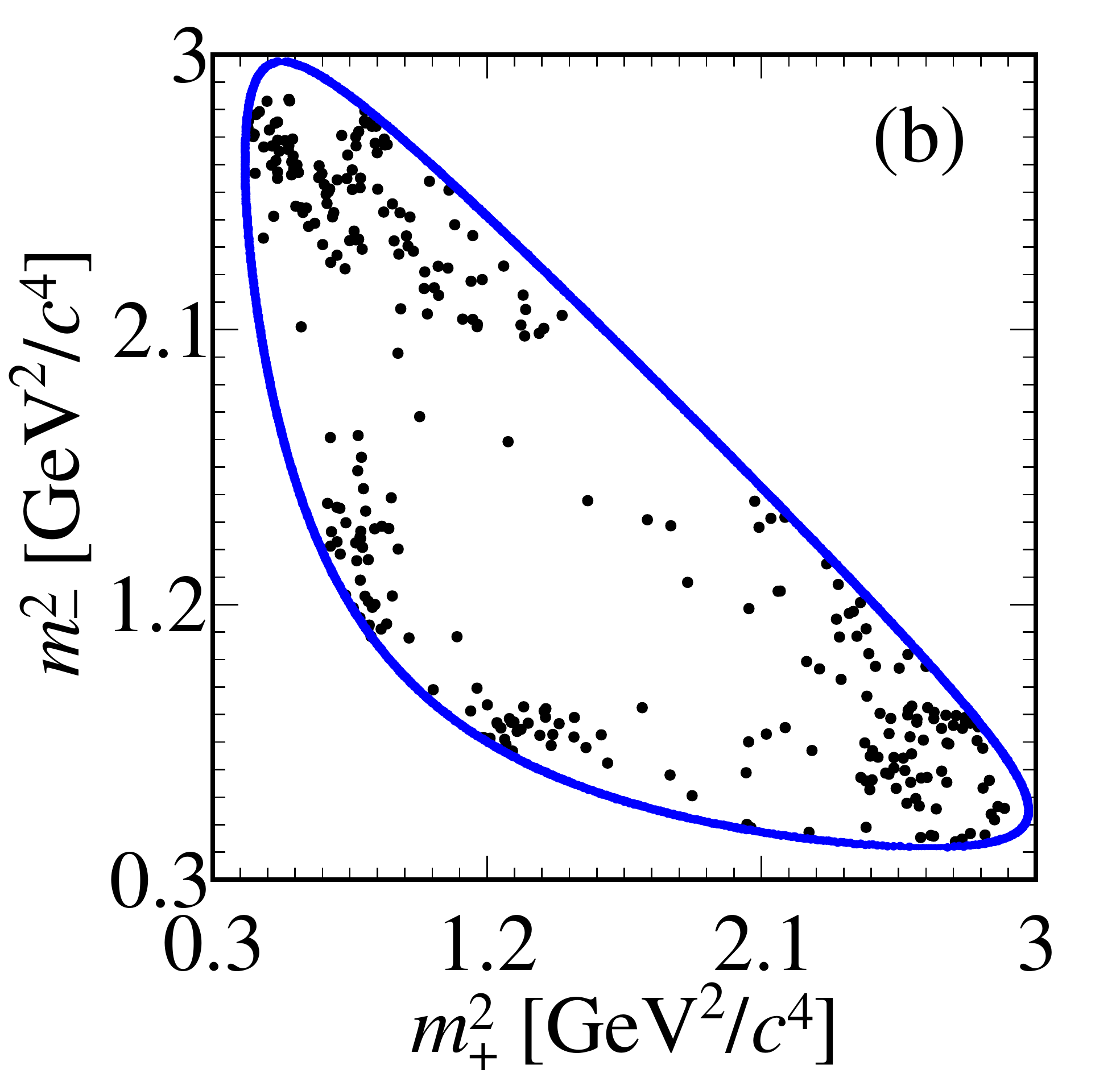} &
\includegraphics[width=0.32\columnwidth]{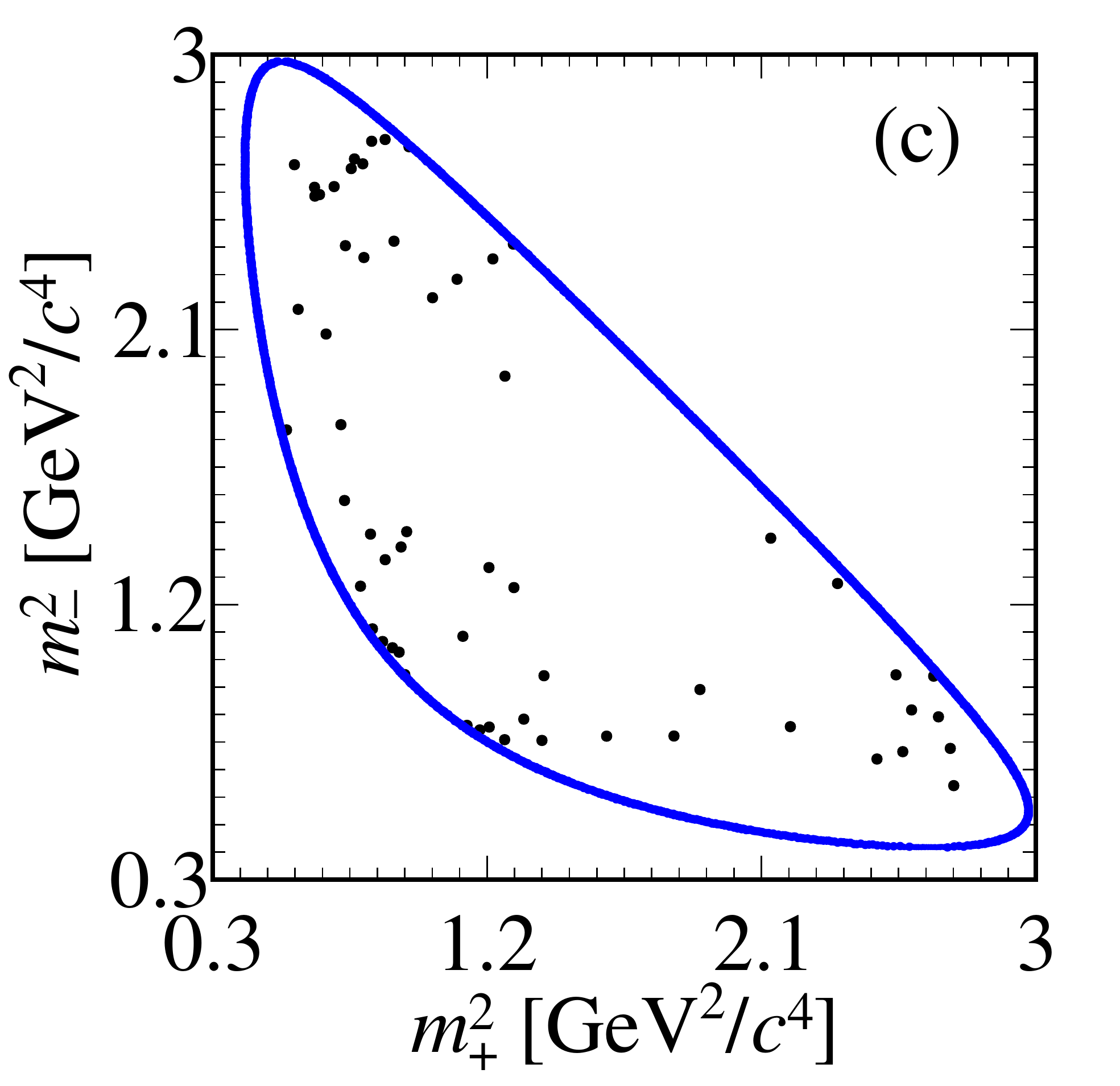}  \\
\includegraphics[width=0.32\columnwidth]{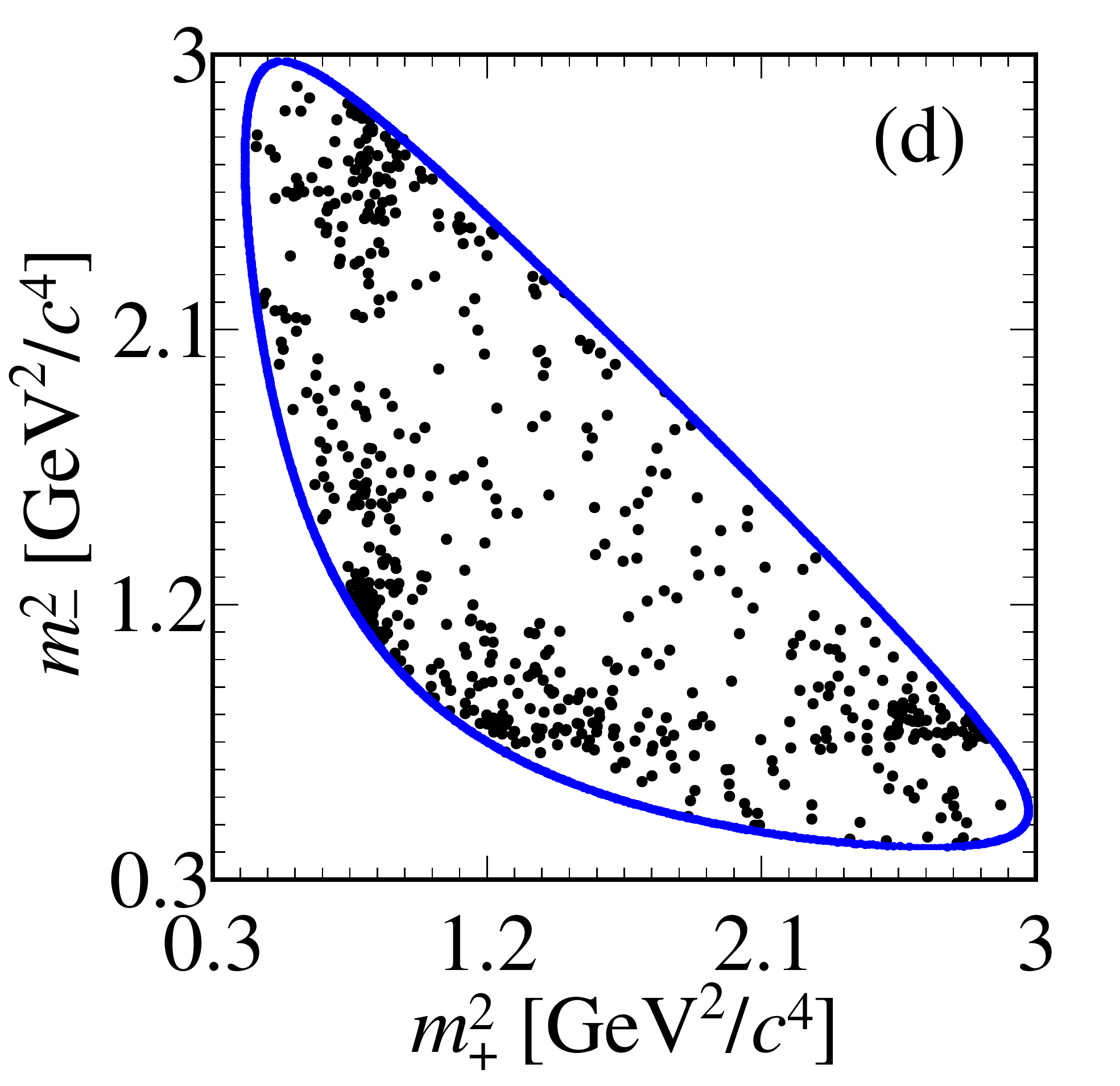} &
\includegraphics[width=0.32\columnwidth]{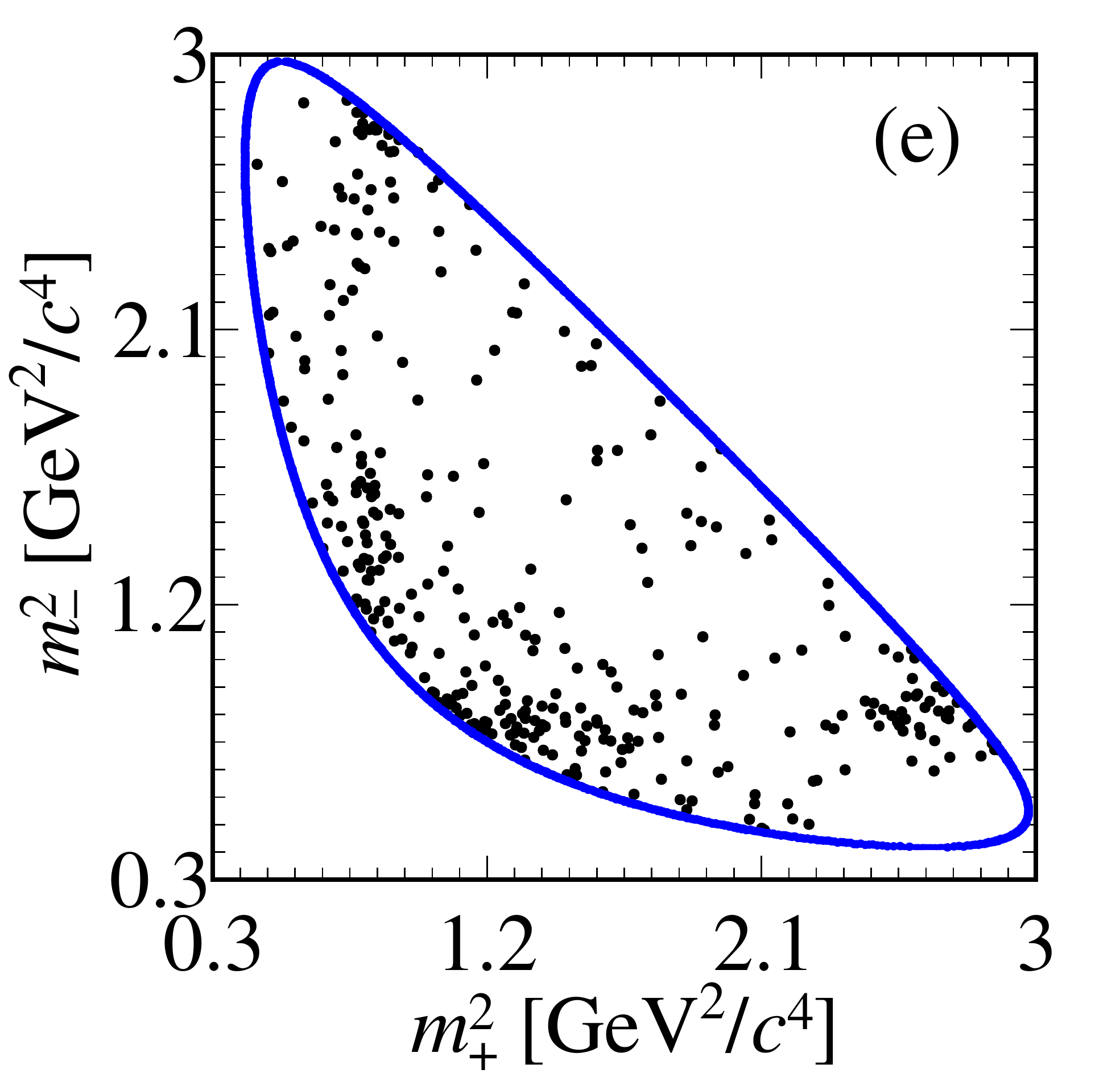} &
\includegraphics[width=0.32\columnwidth]{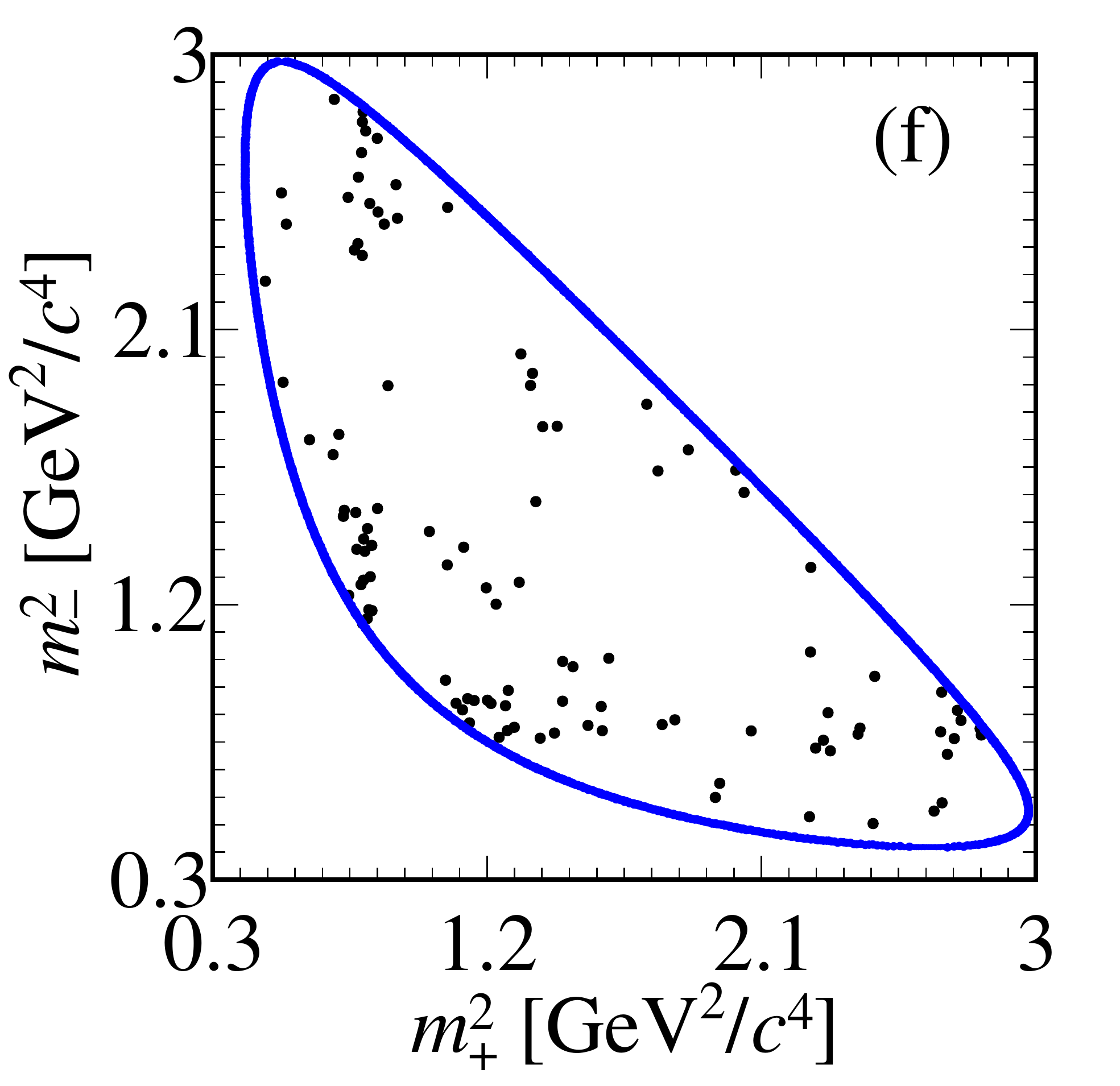}
\end{tabular}
\caption{Dalitz-plot distributions for $D \to \Kspipi$ reconstructed against (a) $D \to \pipipipi$, (b) $D \to \pipipiz$ and  (c) $D \to \KKpiz$, 
and  $D \to \Klpipi$ reconstructed against (d) $D \to \pipipipi$, (e) $D \to \pipipiz$ and  (f) $D \to \KKpiz$. The axis labels $m_\pm^2$ are the invariant-mass squared of the $\pi^\pm K^0_{\rm S, L}$ pair. }  
\label{fig:DPK0pipi}
\end{center}
\end{figure}

Figures~\ref{fig:pipipipi_cp}~(a) and~(b) show the average \Mbc distributions for \CP-tagged $\D\to\pipipipi$ candidates, summed over all tag modes that are \CP-even and \CP-odd eigenstates, respectively, where the \CP-tag final state does not contain a \Kl meson.
Figure~\ref{fig:sig_kspipi} shows the average \Mbc distributions for  $\D\to\pipipipi$, $\D\to\pipipiz$ and $\D\to\KKpiz$  candidates tagged with $\D\to\Kspipi$ decays,
while Figs.~\ref{fig:DPK0pipi}~(a)--(c) show the Dalitz-plot distributions of the tag decay for these three signal modes.

\begin{figure}[!h]
\begin{center}
\begin{tabular}{ccc}
\includegraphics[width=0.32\columnwidth]{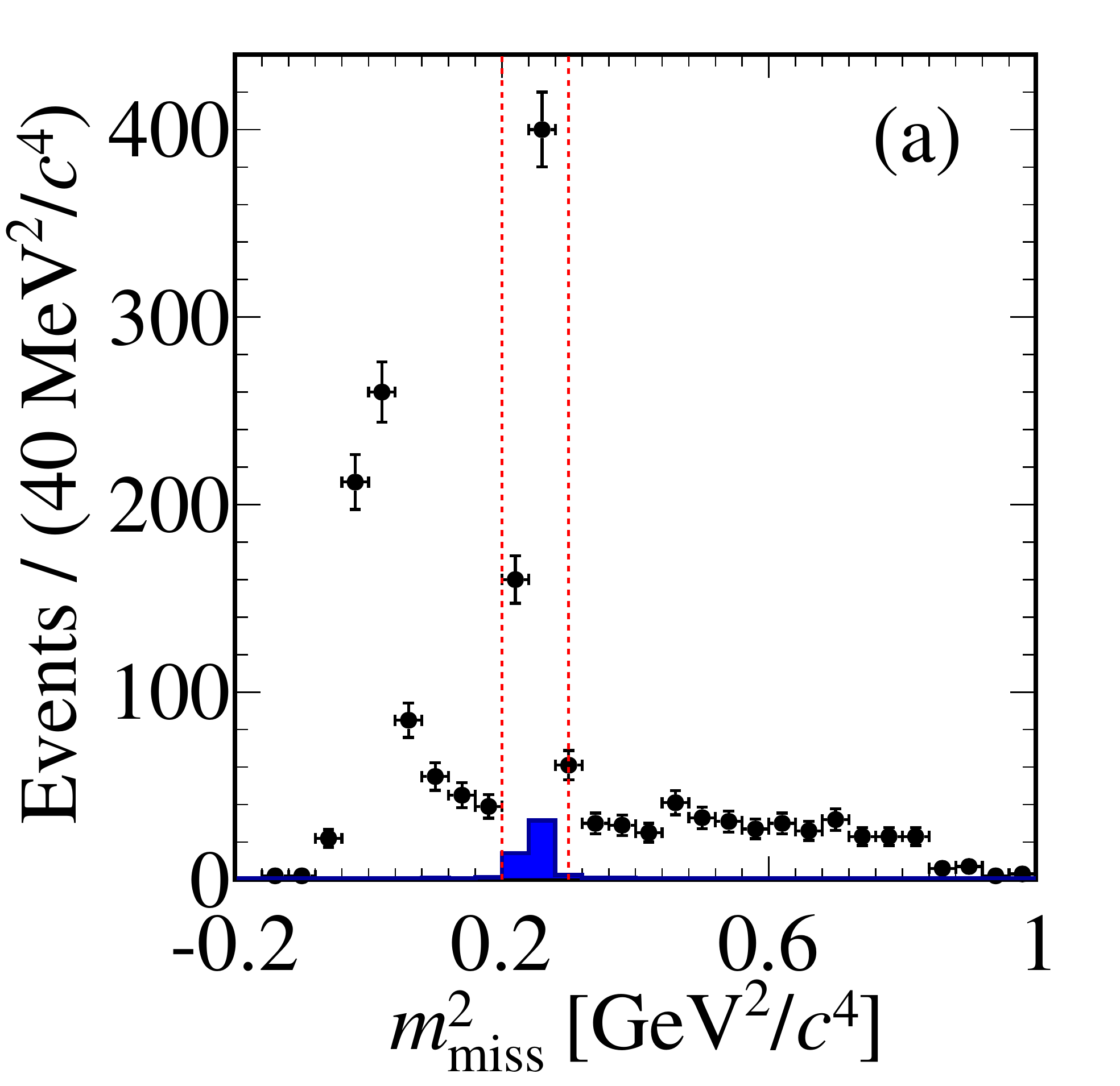} &
\includegraphics[width=0.32\columnwidth]{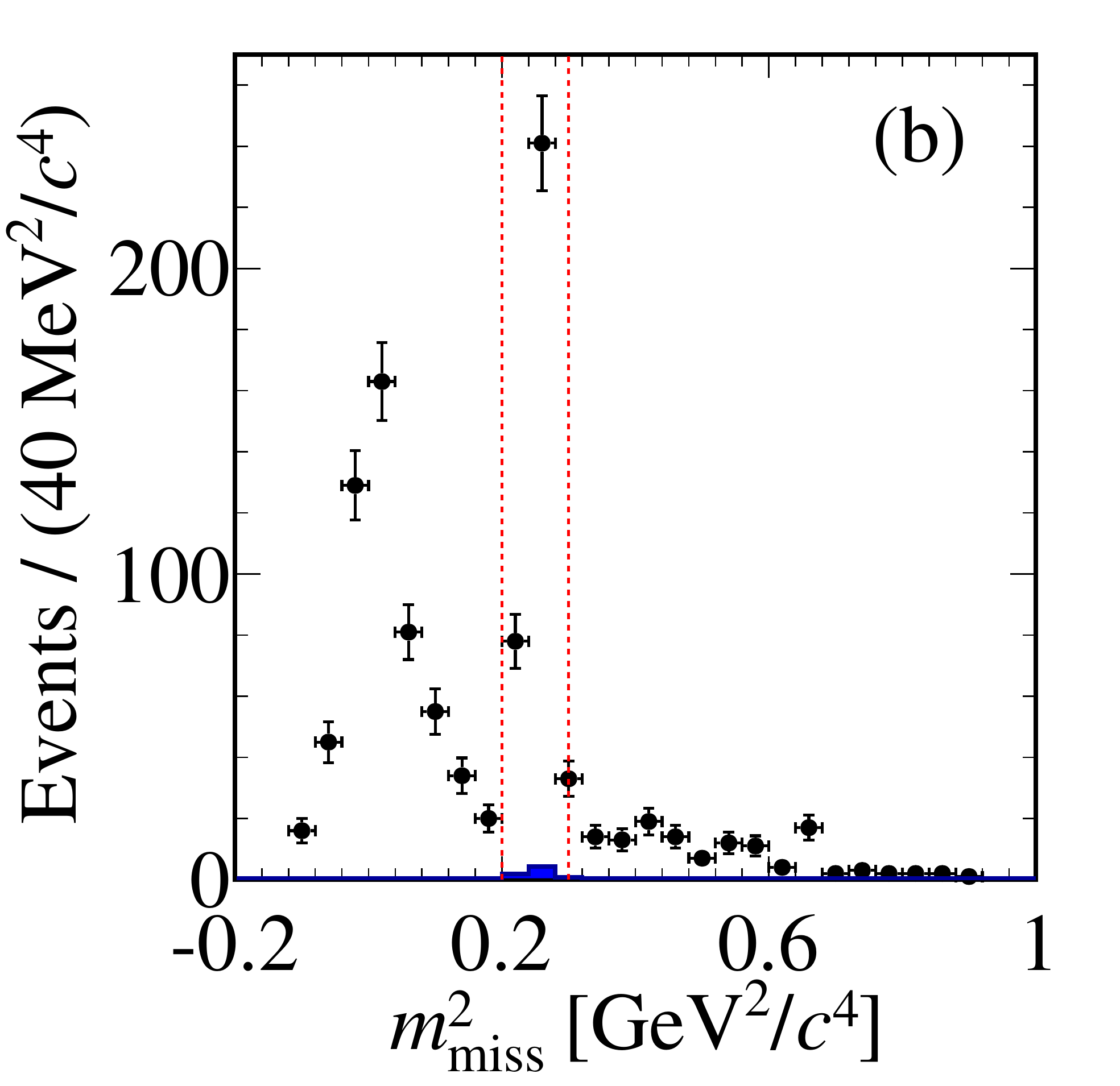} &
\includegraphics[width=0.32\columnwidth]{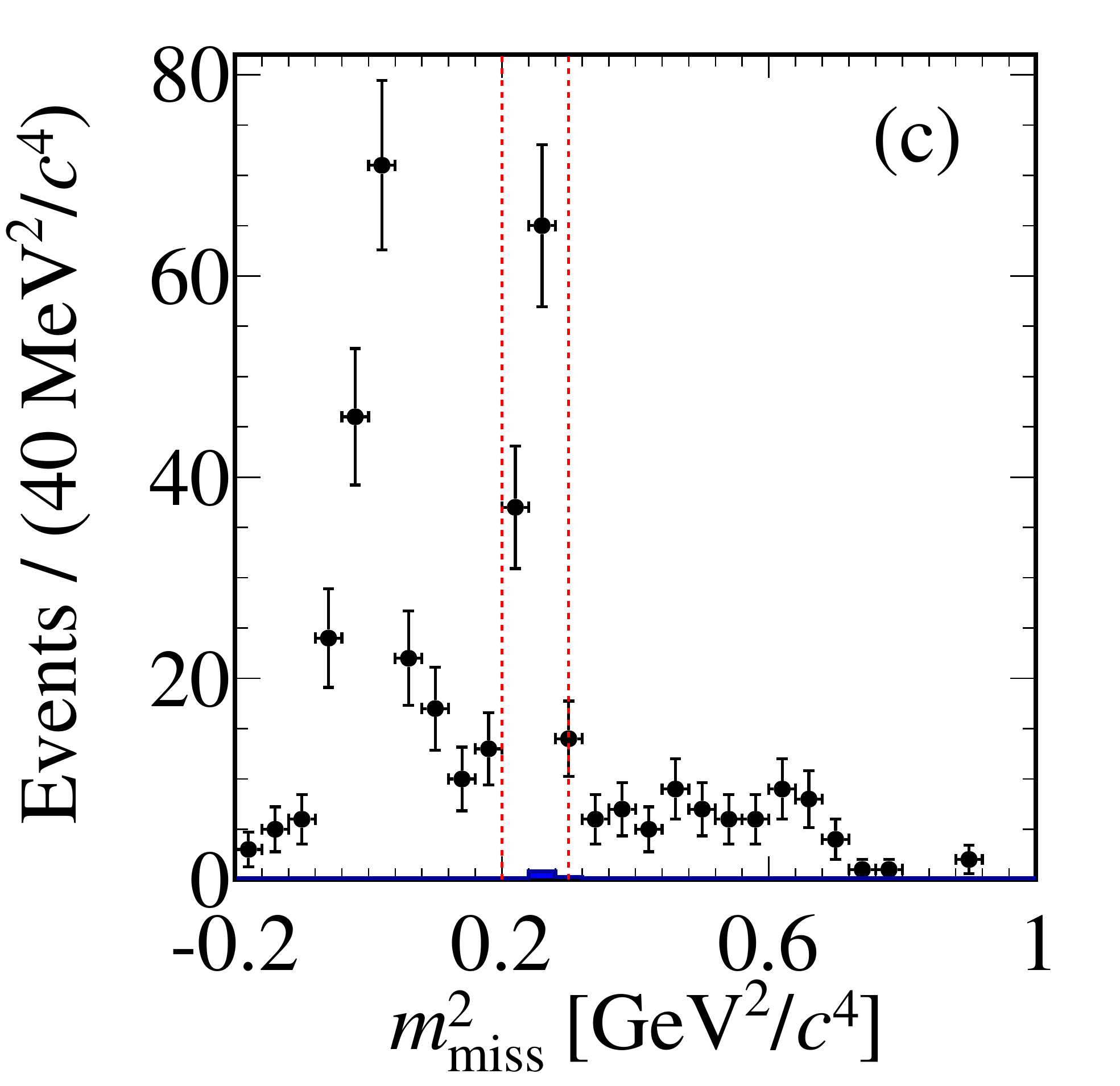} 
\end{tabular}
\caption{\mmsq distributions for (a) $\D\to\pipipipi$, (b) $\D\to\pipipiz$  and (c) $\D\to\KKpiz$ candidates tagged by a $\D\to\Klpipi$ decay. The shaded histogram is the estimated peaking background and the vertical dotted lines indicate the signal region.}  \label{fig:sig_klpipi}
\end{center}
\end{figure}

Many \Kl mesons  do not deposit any reconstructible signal in the detector. However, double-tag candidates can be fully reconstructed using a missing-mass squared (\mmsq) technique \cite{K0LPRL} for tags containing a single \Kl meson. Yields are determined from the signal and sideband regions of the
\mmsq distribution. Figure~\ref{fig:pipipipi_cp}~(c)  shows the \mmsq distributions for $\D\to\pipipipi$ candidates tagged with either a \Klpiz or \Klomega decay.   Figure~\ref{fig:sig_klpipi} shows the  \mmsq distributions for  $\D\to\pipipipi$, $\D\to\pipipiz$ and $\D\to\KKpiz$  candidates tagged with $\D\to\Klpipi$ decays, and Figs.~\ref{fig:DPK0pipi}~(d)--(f) show the corresponding tag-side Dalitz-plot distributions.

In events where more than one pair of decays  is reconstructed an algorithm is applied to select a single double-tag candidate based on the information provided by the \Mbc and \DeltaE variables.  The particular choice of metric varies depending on the category of double tag and is optimised through simulation studies.

The peaking background estimates are determined from the generic Monte Carlo sample of $\Dz\Dzb$ events. 
For  double tags involving a \CP mode without a \Kl meson the peaking backgrounds are found to constitute 5-10\% of the selected events, and are predominantly from residual $D \to K^0_{\rm S} \pi^+\pi^-$ contamination.
The peaking backgrounds for final states with a \Kl are generally larger; for \Klpiz and \Klomega this contamination amounts to 15--20\% of the signal yield, whereas for \Klpipi it is $\sim 10\%$ of the signal yield. The dominant source of peaking background in each case is the equivalent decay containing a \Ks instead of a \Kl meson.   The contamination from specific modes in the other categories of double tags is typically 10\% or less.
The statistical uncertainties on these background estimates arising from the finite size of the simulated samples are included in the total statistical uncertainties on the signal yields.  


The measured double-tag event yields after background subtraction are given in Table~\ref{tab:doubletags}.   
\begin{table}[thb]
\begin{center}
\caption{Double-tagged signal yields after background subtraction.  Information on the entries marked `$\dagger$', not studied in the current analysis, can be found in Ref.~\cite{MINAKSHI}.}\label{tab:doubletags} \vspace*{0.1cm}
\begin{tabular}{l r@{\ $\pm$\ }l r@{\ $\pm$\ }l r@{\ $\pm$\ }l}
\hline\hline 
          & \multicolumn{2}{c}{\pipipipi}  & \multicolumn{2}{c}{\pipipiz} &   \multicolumn{2}{c}{\KKpiz}  \\ 
\hline
\KK       &  $19.3$ &  $6.3$ & \multicolumn{2}{c}{$\dagger$} & \multicolumn{2}{c}{$\dagger$} \\
\pipi     &   $3.3$ &  $8.2$ & \multicolumn{2}{c}{$\dagger$} & \multicolumn{2}{c}{$\dagger$} \\
\Kspizpiz &  $18.6$ &  $5.2$ & \multicolumn{2}{c}{$\dagger$} & \multicolumn{2}{c}{$\dagger$} \\
\Klpiz    &  $49.2$ & $10.9$ & \multicolumn{2}{c}{$\dagger$} & \multicolumn{2}{c}{$\dagger$} \\
\Klomega  &  $22.0$ &  $6.5$ & \multicolumn{2}{c}{$\dagger$} & \multicolumn{2}{c}{$\dagger$} \\
\hline
\Kspiz    & $112.8$ & $11.0$ & \multicolumn{2}{c}{$\dagger$} & \multicolumn{2}{c}{$\dagger$} \\
\Ksomega  &  $41.0$ &  $6.8$ & \multicolumn{2}{c}{$\dagger$} & \multicolumn{2}{c}{$\dagger$} \\
\Ksetagg  &  $18.8$ &  $4.5$ & \multicolumn{2}{c}{$\dagger$} & \multicolumn{2}{c}{$\dagger$} \\
\Ksetahad &   $3.1$ &  $2.7$ & \multicolumn{2}{c}{$\dagger$} & \multicolumn{2}{c}{$\dagger$} \\
\Ksetap   &   $9.3$ &  $3.3$ & \multicolumn{2}{c}{$\dagger$} & \multicolumn{2}{c}{$\dagger$} \\
\hline
\Kspipi   & $217.9$ & $16.8$ & $289.2$ & $17.6$ & $52.5$ & $7.8$ \\
\Klpipi   & $485.0$ & $26.3$ & $291.1$ & $19.2$ & $78.1$ & $11.1$ \\
\hline
\pipipipi  &  $41.0$     &  $16.3$        & $75.5$ & $15.7$ & \multicolumn{2}{c}{---} \\
\hline\hline   
\end{tabular}
\end{center}
\end{table}

 Knowledge of the single-tag yields of the \CP-eigenstate modes is required for normalisation purposes.  Since the single-tag reconstruction criteria applied are identical to those employed in Ref.~\cite{MINAKSHI}, all information on these yields is taken from the earlier publication.  It is also necessary to know the single-tag yield for the decay $D \to \pi^+\pi^-\pi^0$.  A fit to the \Mbc distribution returns a result of $29998 \pm 320$ signal candidates, after the subtraction of small peaking-background contributions.

\section{Analysis with the \CP tags}
\label{sec:cptags}

The yields of the single and double tags containing a \CP eigenstate  are used as inputs to determine the \CP-even fraction, \Ffourpi.
 Following on from Eq.~\ref{eq:Ndoubletag}, the expected number of observed events, $M$, where one $D$ meson decays to the $\pipipipi$ final state, and the other decays to  $X$, a \CP eigenstate with eigenvalue $\eta_{\CP}$, is given by 
\begin{equation}
M(4\pi | X ) = 2\NDDbar\,\mathcal{B}(4\pi)\mathcal{B}(X)\varepsilon(4\pi | X ) \left[ 1- \eta_{\CP}\Big(2\Ffourpi -1\Big)\right],
\end{equation}
\noindent where \NDDbar is the number of \D\Db pairs, $\mathcal{B}(4\pi)$ and $\mathcal{B}(X)$ are the branching fractions for the two reconstructed final states and $\varepsilon(4\pi | X )$ is the efficiency of reconstructing such a double tag. The double tag yield is denoted by $M^-$ ($M^+$) for \CP-even (\CP-odd) tags.  
Experimentally it is advantageous to eliminate dependence on \NDDbar, the branching fractions and the reconstruction efficiency, which can be achieved by normalising by the single-tag yields. The yield of single tags, $S^+$ ($S^-$) decaying to a \CP-odd (\CP-even) eigenstate $X$, is given by 
\begin{equation}
S(X) =  2 \NDDbar \, \mathcal{B}(X)\varepsilon(X),
\label{eq:singtag}
\end{equation} 
where $\varepsilon(X)$ is the reconstruction efficiency of the single tag.
The small effects of \Dz\Dzb mixing are eliminated from the measurement by correcting the measured single-tag yields $S^{\pm}_{\rm meas}$ such that $S^{\pm}=S^{\pm}_{\rm meas}/(1-\eta_{\CP} y_D)$ where $y_D = (0.62 \pm 0.08) \%$ is the well-known $D$-mixing parameter~\cite{HFAG}. 
A further correction is applied in the case of the tags $K^+K^-$ and $\pi^+\pi^-$ because of  the differing selection requirements for the single and double-tag case, as described in Sect.~\ref{sec:eventsel}. This correction factor is determined by taking the ratio of the selection efficiency of the single tag from simulation with the two differing selections. It is determined to be 1.15 and 1.10 for the $D \to \Kp\Km$ and $D \to \pip\pim$ modes, respectively, with an uncertainty of $\pm 0.05$.  The other uncertainties on these single-tag yields are assigned following the same procedure described in Ref.~\cite{MINAKSHI}. 

For the case of the two \CP tags involving a $K^0_{\rm L}$ meson a different treatment is required, since it is not possible to measure the single-tag yield directly for these modes. Following the procedure described in Ref.~\cite{MINAKSHI}, the effective single-tag yield is evaluated using Eq.~\ref{eq:singtag}, where the effective single-tag efficiency $\varepsilon(K_{\rm L}^0X)$ is calculated from the ratio of $\varepsilon(4\pi|K_{\rm L}^0X)/\varepsilon(4\pi)$, and the  leading systematic uncertainties are associated with  the  branching fractions and the value used for the effective single-tag efficiency.  The effective single-tag yields are determined to be 21726 $\pm$ 3497 and 9124 $\pm$ 4105 for $K_{\rm L}^0\pi^0$ and $K_{\rm L}^0\omega$, respectively.

\begin{figure}[tb]
\begin{center}
\begin{tabular}{cc}
\includegraphics[width=0.45\columnwidth]{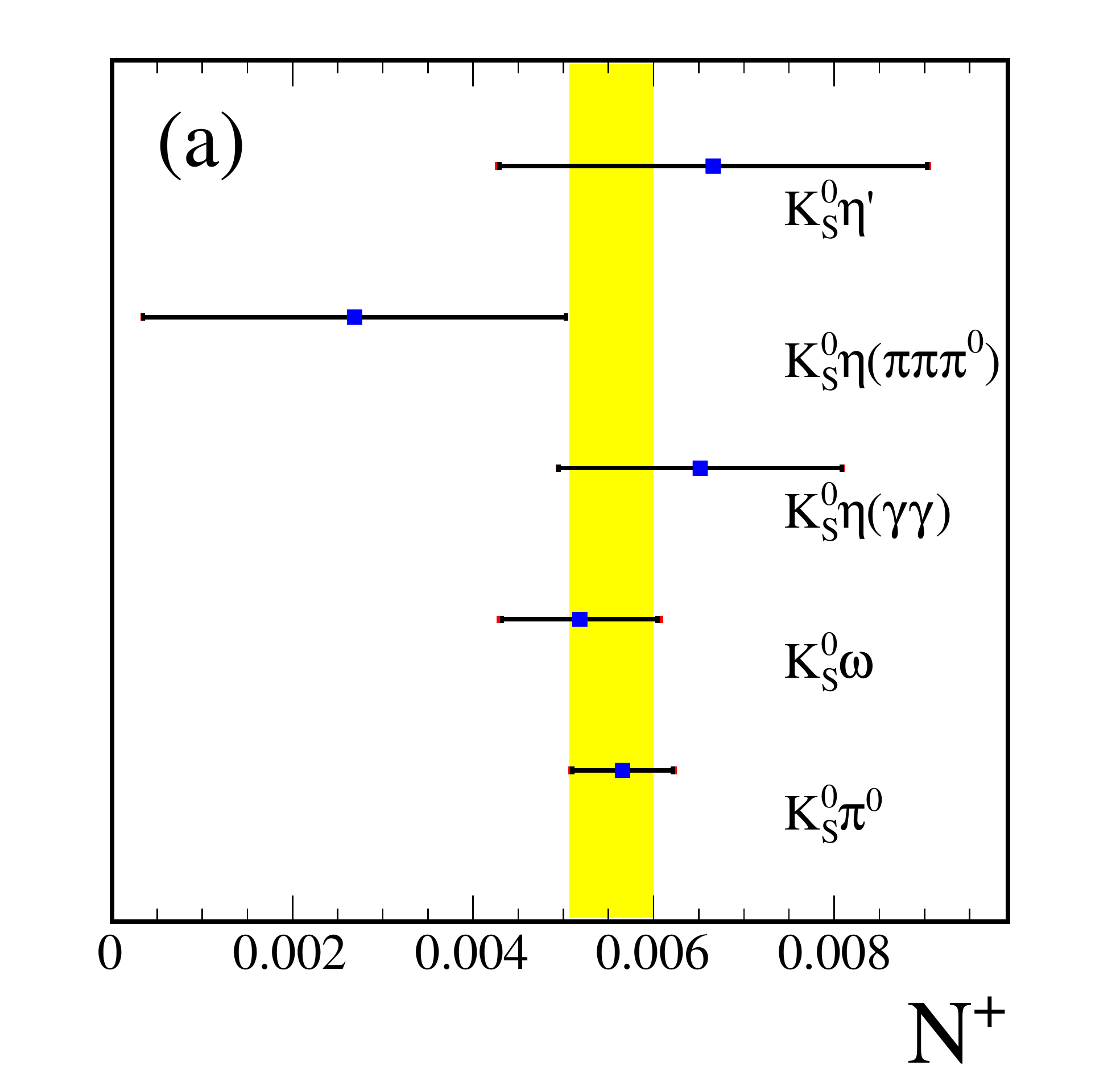} &
\includegraphics[width=0.45\columnwidth]{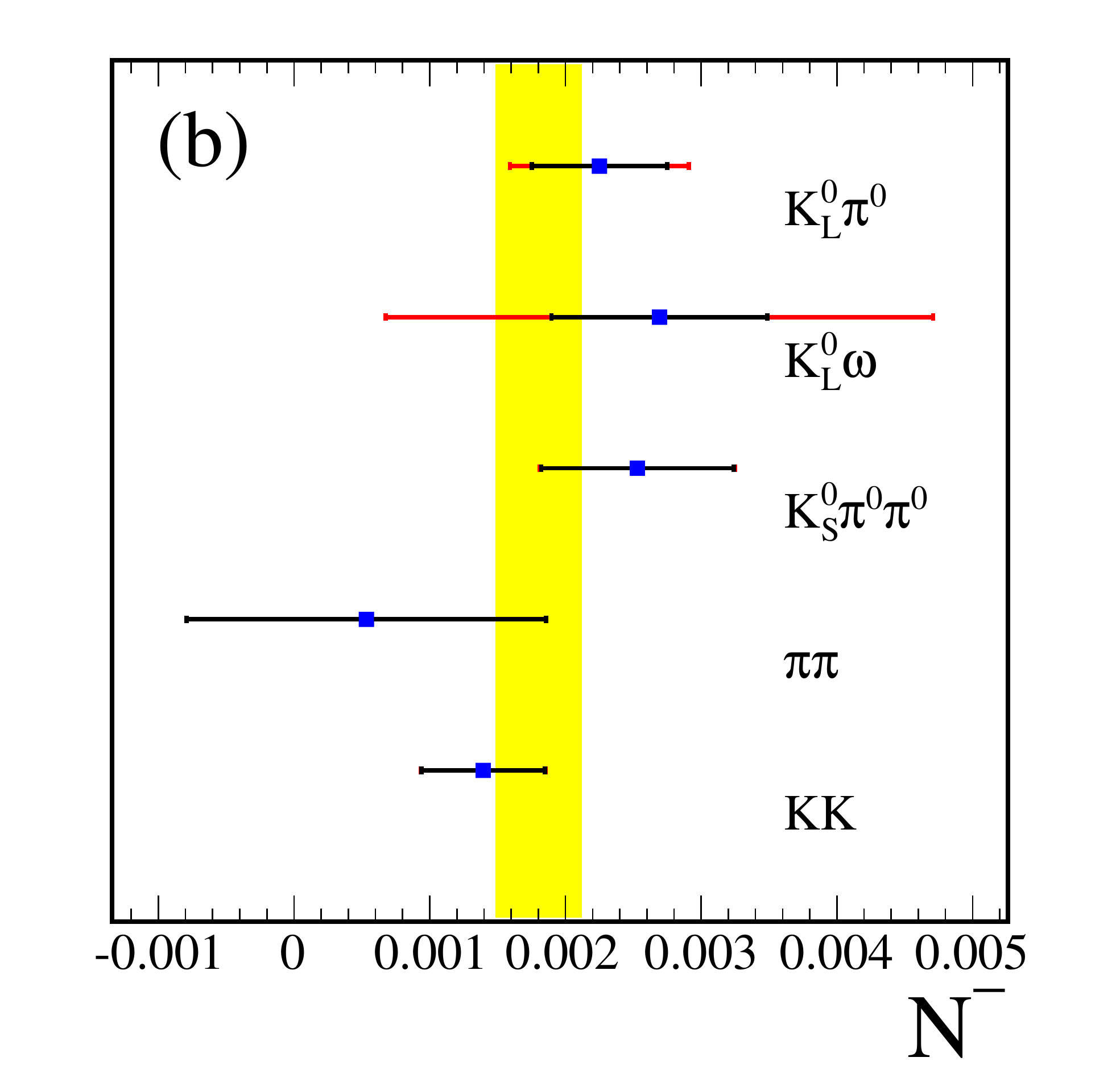} 
\end{tabular}
\caption{$D \to \pi^+\pi^-\pi^+\pi^-$ results for (a) $N^+$  and (b) $N^-$.
In each plot the vertical yellow band indicates the value obtained from the combination of all tags. The black portion of the uncertainty represents the statistical uncertainty only while the red represents the total.  }  \label{fig:results}
\end{center}
\end{figure}

 Assuming that the reconstruction efficiencies of each $D$ meson are independent, then the ratio of the double-tagged and single-tagged yields are independent of the branching fraction and reconstruction efficiency of the \CP tag and \NDDbar. This ratio is defined as $N^+ \equiv M^+/S^+$, with an analogous expression for $N^-$. The \CP-even fraction \Ffourpi is then given by
\begin{equation}
\Ffourpi = \frac{N^+}{N^+ + N^-}.
\end{equation}
The measured values for $N^+$ and $N^-$ for each \CP tag are displayed in Fig.~\ref{fig:results}.  It can be seen that there is consistency between the individual tags for each measurement. The mean value $<N^+> = (5.54 \pm 0.46) \times 10^{-3}$ is significantly larger than $<N^-> = (1.80 \pm 0.32) \times 10^{-3}$, indicating that the \pipipipi final state is predominantly \CP even.

If the acceptance across the phase space of the $D \to \pipipipi$ decay is not uniform it has the potential to bias the measurement of \Ffourpi. 
Using simulated data the selection efficiency of individual pions in  $D \to \pipipipi$ decays  is determined in bins of  momentum and polar angle with respect to the beam direction.
The candidates in data are then weighted by the normalised efficiency. 
Each pion is treated independently and an overall weight, typically lying within 5--10\% of unity,  is found by multiplying  the individual weights. 
The scaled signal yields are used to re-determine \Ffourpi and the difference between this and the value found without efficiency correction is 0.008, which is taken as the systematic uncertainty due to non-uniform acceptance.

Using the \CP tags only, and accounting for the correlations between the systematic uncertainties, yields $F_+^{4\pi} = 0.754 \pm 0.031 \pm 0.021$, where the first uncertainty is statistical and the second systematic.

\section{Analysis with the $\D\to\Kslpipi$ tags}
\label{sec:k0pipitags}

For each of the signal samples that are tagged by $D \to \Kspipi$ or $D \to \Klpipi$ decays the Dalitz plot of the tag mode is divided into eight pairs of symmetric bins by the line $m_+^2 = m_-^2$, where
$m_\pm^2$ is the invariant-mass squared of the $\pi^\pm K^0_{\rm S,L}$ pair. The bins lying on one side of this line ($m_+^2 > m_-^2$) are labelled $-1 \to -8$, and those on the other side $1 \to 8$. The binning definition follows the `Equal $\Delta \delta_D$ BABAR 2008'  scheme of Ref.~\cite{CLEOKSPIPI}, in which the boundaries are chosen according to the strong-phase prediction of a model developed by the BaBar collaboration~\cite{BABAR_2008}.  The expected distribution of entries is symmetric and so the analysis considers the absolute bin number $|i|$, which contains the contents of the pair of bins $-i$ and $i$.   

Following Eq.~\ref{eq:bins},
the expected population of  bin $|i|$ for signal decays with $K^0_{\rm S}\pi^+\pi^-$ tags is
\begin{equation}
M_{|i|} = h \left[ K_i + K_{-i} - \Big(2\Fp - 1\Big) 2c_i \sqrt{K_i K_{-i}}\right],
\label{eq:kspipi}
\end{equation}
where $h$ is a normalisation factor specific to the signal category, $K_i$ is the flavour-tagged fraction, being the proportion of $K^0_{\rm S} \pi^+\pi^-$ decays to fall in bin $i$ in the case that the mother particle  is known to be a $D^0$ meson, and  $c_i$ is the cosine of the strong-phase difference between $D^0$ and $\bar{D}^0$ decays averaged in bin $i$ and weighted by the absolute decay rate  (a precise definition may be found in Ref.~\cite{ANTONBONDAR}). The only difference between the form of this expression and the case when the signal decays into a pure $CP$-even eigenstate~\cite{ANTONBONDAR} is the prefactor of $(2F_+ - 1)$ in the final term.

Similarly, when the tagging meson decays to \Klpipi then the number of double-tag decays produced in bin $|i|$ is 
\begin{equation}
M'_{|i|} = h' \left[ K_i' + K_{-i}' + \Big(2\Fp - 1\Big)  2c'_i \sqrt{K'_i K'_{-i}}\,\right],
\label{eq:klpipi}
\end{equation}
where the primed quantities are now specific to this case.
The reversed sign in front of the final term reflects the fact that the \Kl meson is almost entirely a \CP-odd eigenstate.

 The values of $c_i$ and $c'_i$ within these bins have been measured by the CLEO collaboration~\cite{CLEOKSPIPI}.   The values of the $K_i$ parameters are taken from an analysis of the predictions of various $B$-factory models~\cite{BABAR_2010,BELLE_2010,BABAR_2005,BABAR_2008} presented in Ref.~\cite{SOL}, and those of the $K'_i$ parameters from measurements performed with CLEO-c data~\cite{BRISBANE}.

The double-tagged samples are analysed to determine the background-subtracted signal yield in each Dalitz-plot bin.
The distribution of background between the different  bins is assigned according to its category.   Flat background is assumed to contribute proportionally to the bin area.
Peaking backgrounds that occur on the signal side affect the  distribution of tag decays in $K^0_{\rm S, L} \pi^+\pi^-$ phase space according to their nature.  For example, in the case of 
$D \to K^0_{\rm S} \pi^0$ decays that are wrongly reconstructed as $D \to \pi^+\pi^-\pi^0$, the tag decay will be in a \CP-even state and distributed accordingly.
Similarly,  the distribution of   $K^0_{\rm S}(\pi^0\pi^0)\pi^+\pi^-$ decays that are misreconstructed as $K^0_{\rm L}\pi^+\pi^-$ tags is well understood and modelled appropriately.   The distribution of the residual $K^0_{\rm S} \pi^+\pi^-$ vs. $K^0_{\rm S} \pi^+\pi^-$ events that contaminate the $\pipipipi$ vs. $K^0_{\rm S} \pi^+\pi^-$ selection is determined from data by inverting the $K^0_{\rm S}$ veto on the signal decay.   

It is also necessary to account for relative bin-to-bin efficiency variations in the background-subtracted signal yields.  The correction factors are determined from simulation and typically differ $\lesssim 5 \%$ from unity. The signal yields in each bin after background subtraction and relative efficiency correction are shown in Table~\ref{tab:kspipi} for \Kspipi tags and in  Table~\ref{tab:klpipi} for \Klpipi tags. 

\begin{table}[thb]
\begin{center}
\caption{Double-tagged signal yields vs.\ \Kspipi after background subtraction in absolute bin numbers of the $\D \to\Kspipi$ Dalitz plot.  
The yields are corrected for relative bin-to-bin efficiency variations and then scaled so that the totals match the values in Table~\ref{tab:doubletags}.}\label{tab:kspipi}\vspace*{0.1cm}
\begin{tabular}{l r@{\ $\pm$\ }l r@{\ $\pm$\ }l r@{\ $\pm$\ }l}
\hline\hline 
$|i|$ & \multicolumn{2}{c}{\pipipipi} & \multicolumn{2}{c}{\pipipiz} & \multicolumn{2}{c}{\KKpiz} \\
\hline
1 & $30.8$ & $7.0$ & $29.9$ & $6.3$ & $12.6$ & $4.1$  \\
2 & $19.8$ & $5.3$ & $19.1$ & $4.8$ & $4.6$ & $2.6$  \\
3 & $16.4$ & $4.5$ & $27.2$ & $5.2$ & $6.9$ & $2.5$  \\
4 & $10.1$ & $3.4$ & $18.5$ & $4.4$ & $1.6$ & $1.5$  \\
5 & $55.1$ & $8.1$ & $96.9$ & $10.0$ & $8.4$ & $3.1$  \\
6 & $21.1$ & $5.1$ & $31.2$ & $5.8$ & $4.4$ & $2.4$  \\
7 & $27.7$ & $6.0$ & $34.6$ & $6.3$ & $7.7$ & $2.8$  \\
8 & $36.9$ & $6.8$ & $31.8$ & $6.0$ & $6.2$ & $2.5$  \\
\hline\hline
\end{tabular}
\end{center}
\end{table}

\begin{table}[thb]
\begin{center}
\caption{Double-tagged signal yields vs.\ \Klpipi  after background subtraction in absolute bin numbers of the $\Dz\to\Klpipi$ Dalitz plot.  
The yields are corrected for relative bin-to-bin efficiency variations.}\label{tab:klpipi} \vspace*{0.1cm}
\begin{tabular}{l r@{\ $\pm$\ }l r@{\ $\pm$\ }l r@{\ $\pm$\ }l}
\hline\hline 
$|i|$ & \multicolumn{2}{c}{\pipipipi } & \multicolumn{2}{c}{\pipipiz} & \multicolumn{2}{c}{\KKpiz} \\
\hline
1 & $134.1$ & $13.9$ & $89.2$ & $11.1$ & $17.3$ & $6.1$ \\
2 & $59.2$ & $8.9$ & $32.9$ & $6.9$ & $8.8$ & $4.0$  \\
3 & $55.4$ & $8.7$ & $31.0$ & $6.3$ & $4.1$ & $3.1$  \\
4 & $20.3$ & $5.8$ & $7.0$ & $3.7$ & $0.1$ & $1.9$  \\
5 & $46.0$ & $8.7$ & $6.7$ & $4.8$ & $2.1$ & $3.1$  \\
6 & $24.6$ & $6.2$ & $14.7$ & $5.0$ & $10.0$ & $3.7$  \\
7 & $61.2$ & $9.0$ & $46.7$ & $7.8$ & $17.6$ & $4.7$  \\
8 & $84.1$ & $10.8$ & $62.9$ & $8.9$ & $18.1$ & $5.1$  \\
\hline\hline
\end{tabular}
\end{center}
\end{table}

\begin{figure}[thb]
\begin{center}
\begin{tabular}{ccc}
\includegraphics[width=0.32\columnwidth]{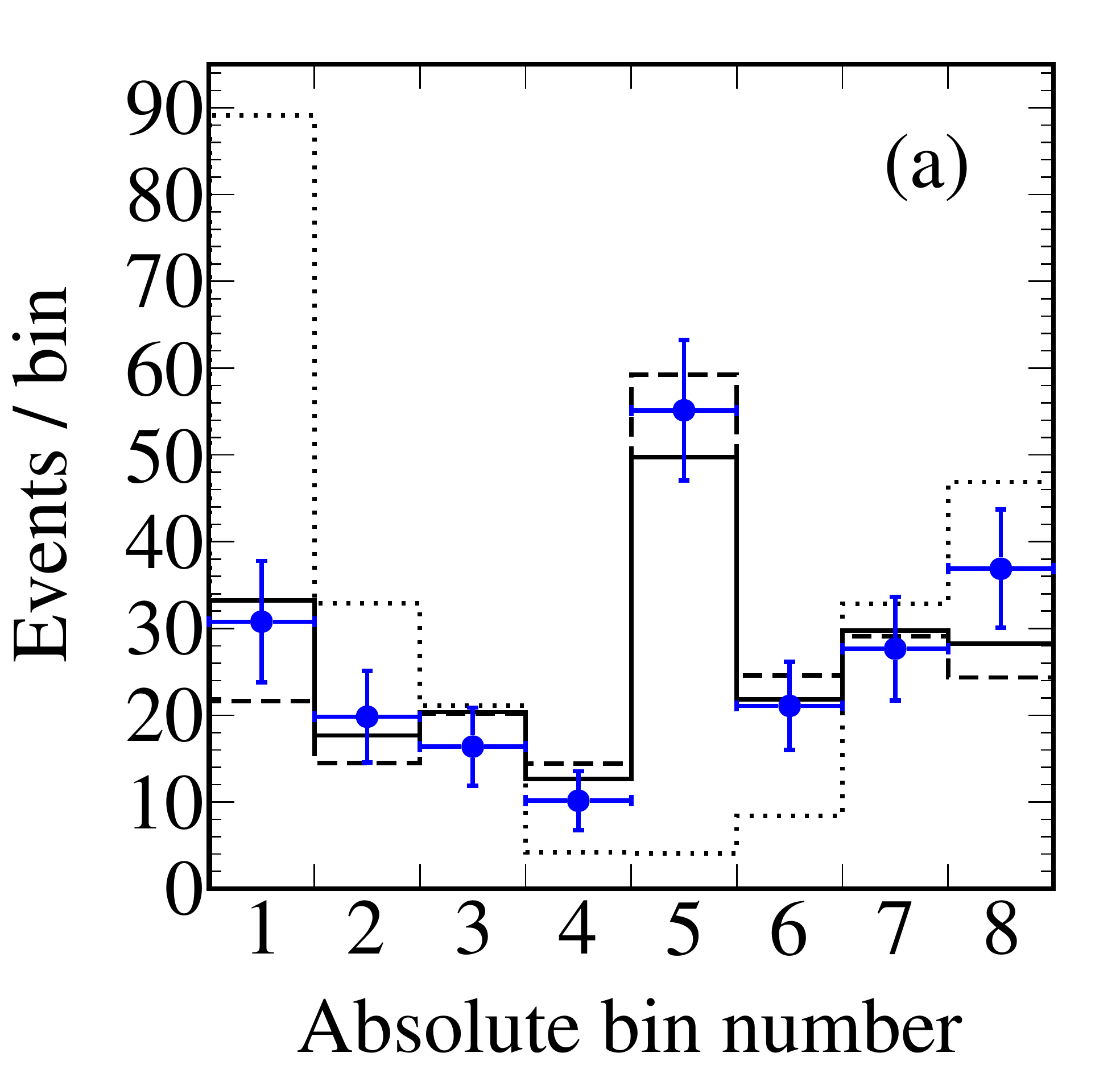} &
\includegraphics[width=0.32\columnwidth]{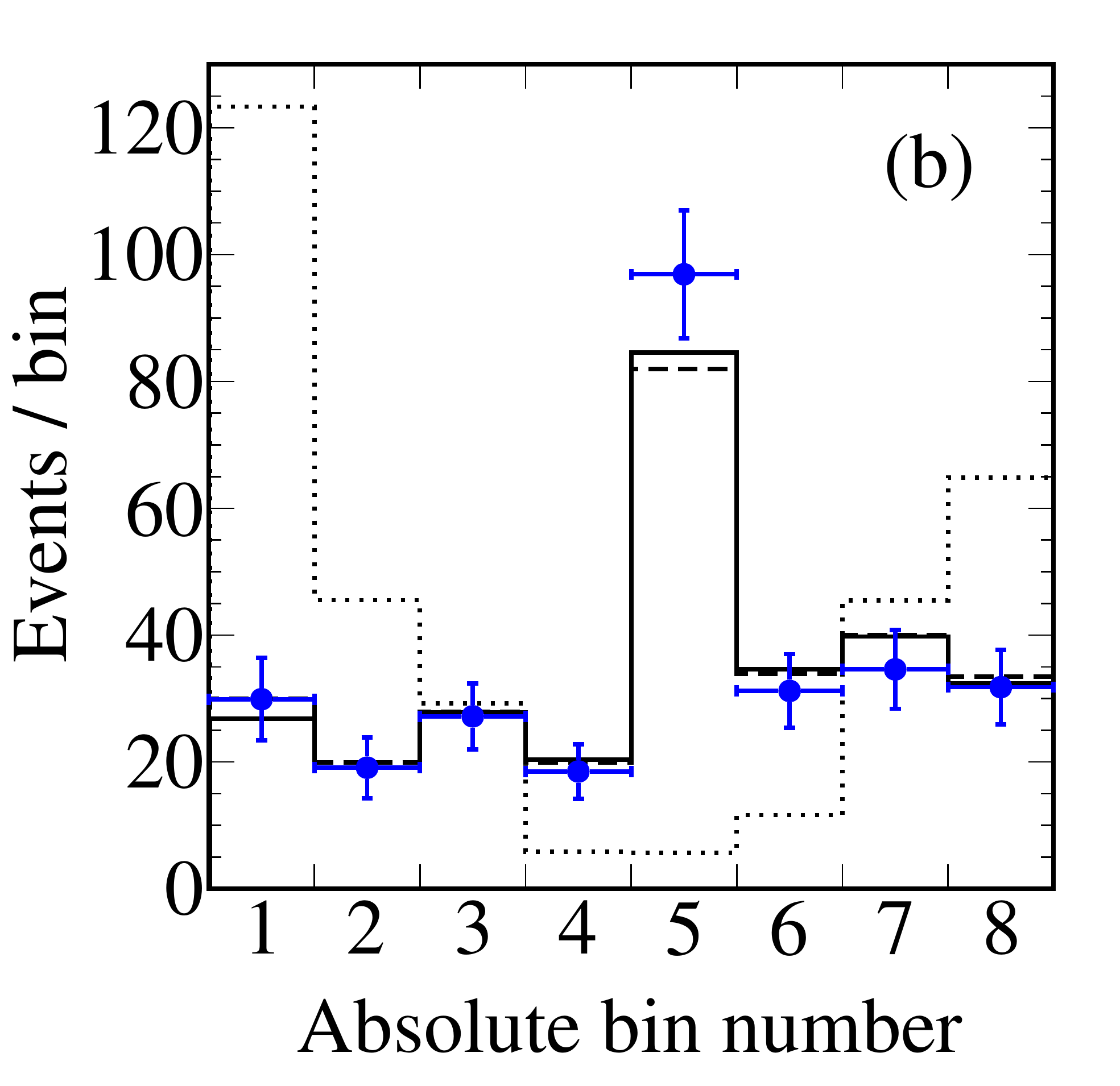} &
\includegraphics[width=0.32\columnwidth]{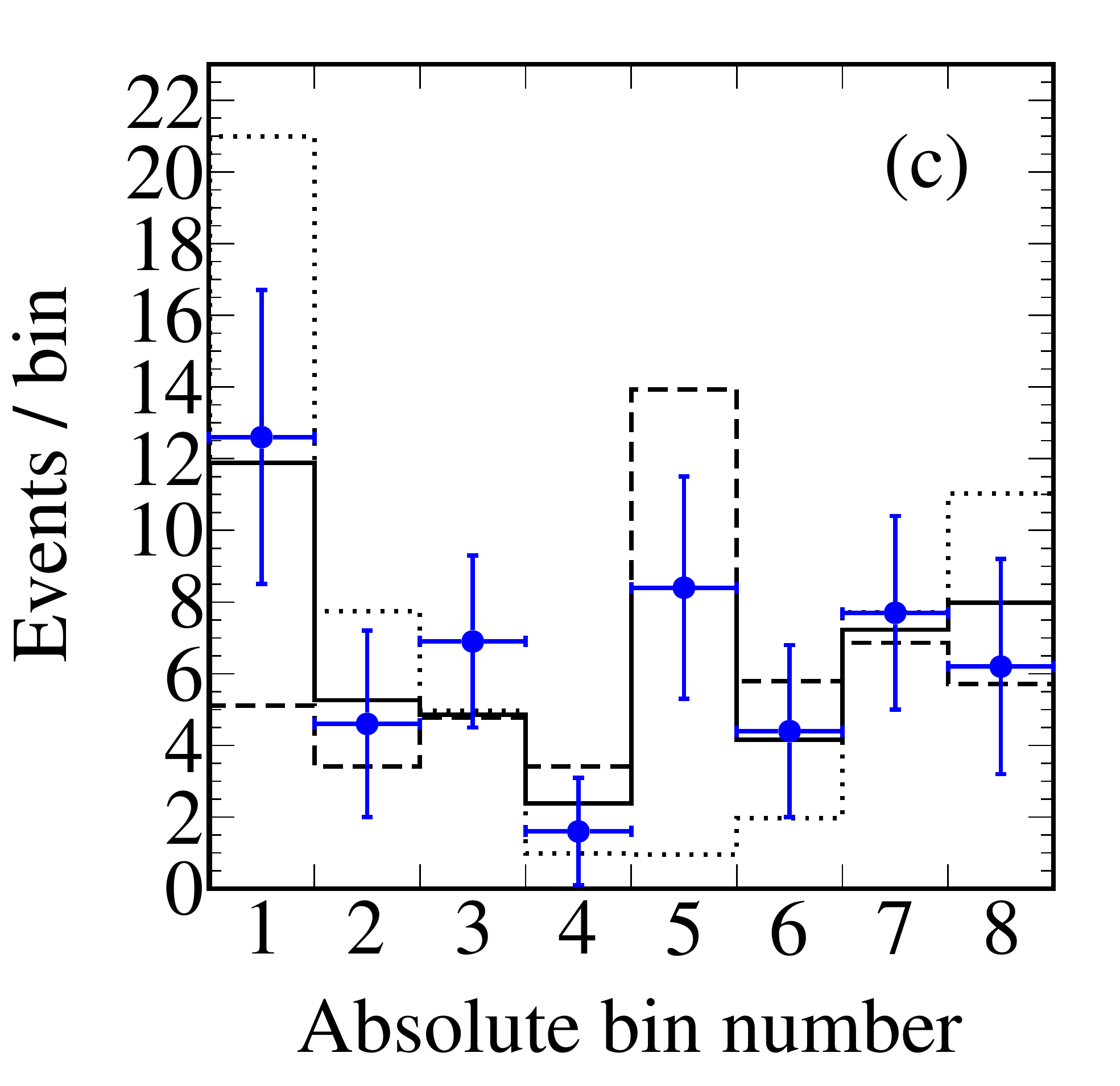} 
\end{tabular}
\caption{Data (points) and fit results (solid line) in absolute bin numbers for \Kspipi tags vs. (a) $D \to \pipipipi$, (b) $D \to 
\pipipiz$ and (c) $D \to \KKpiz$.  Also shown in each case is the expectation if $\Fp = 0$ (dotted line) or $\Fp = 1$ (dashed line). }  
\label{fig:fitkspipi}
\end{center}
\end{figure}

\begin{figure}[!h]
\begin{center}
\begin{tabular}{ccc}
\includegraphics[width=0.32\columnwidth]{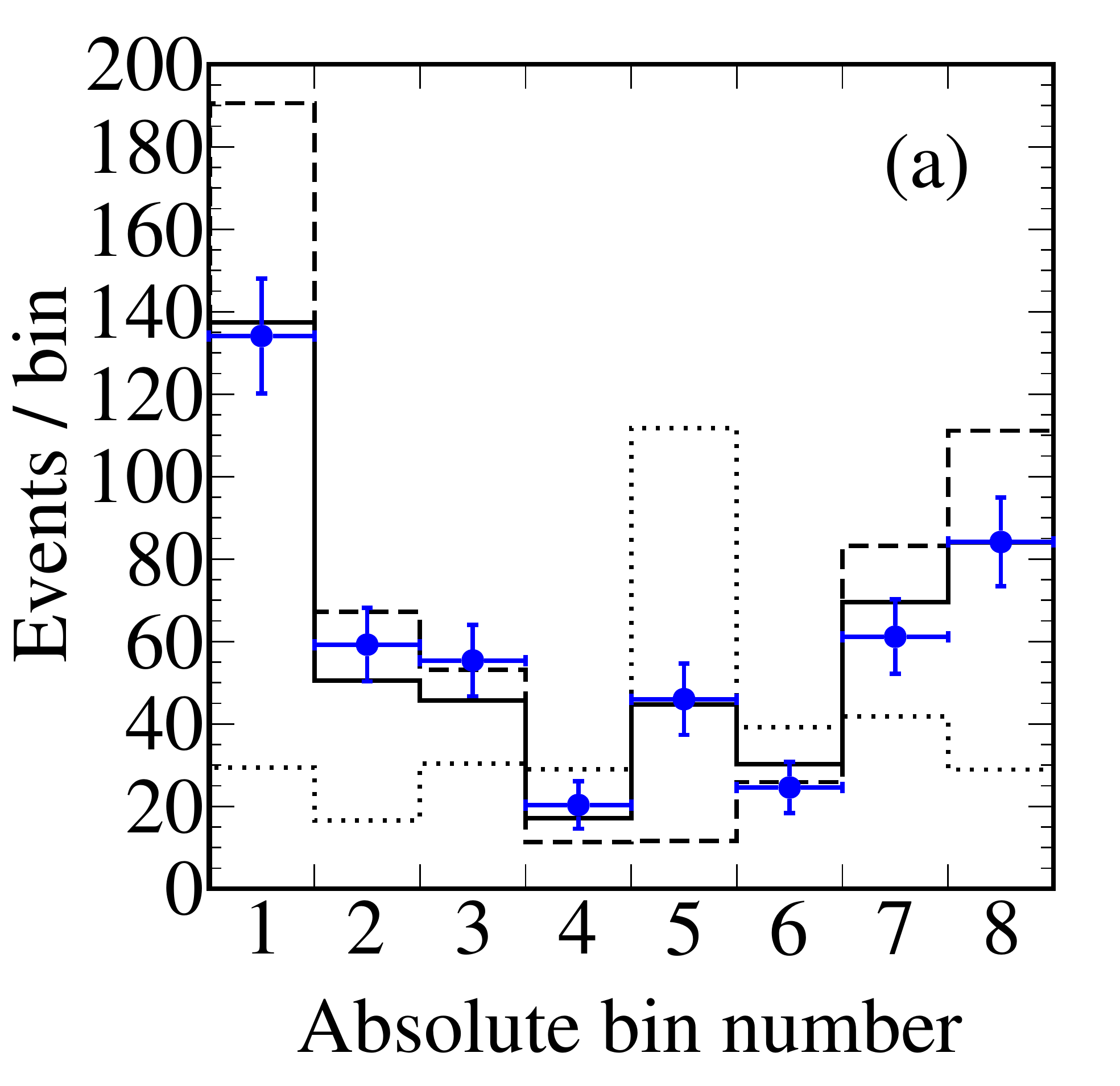} &
\includegraphics[width=0.32\columnwidth]{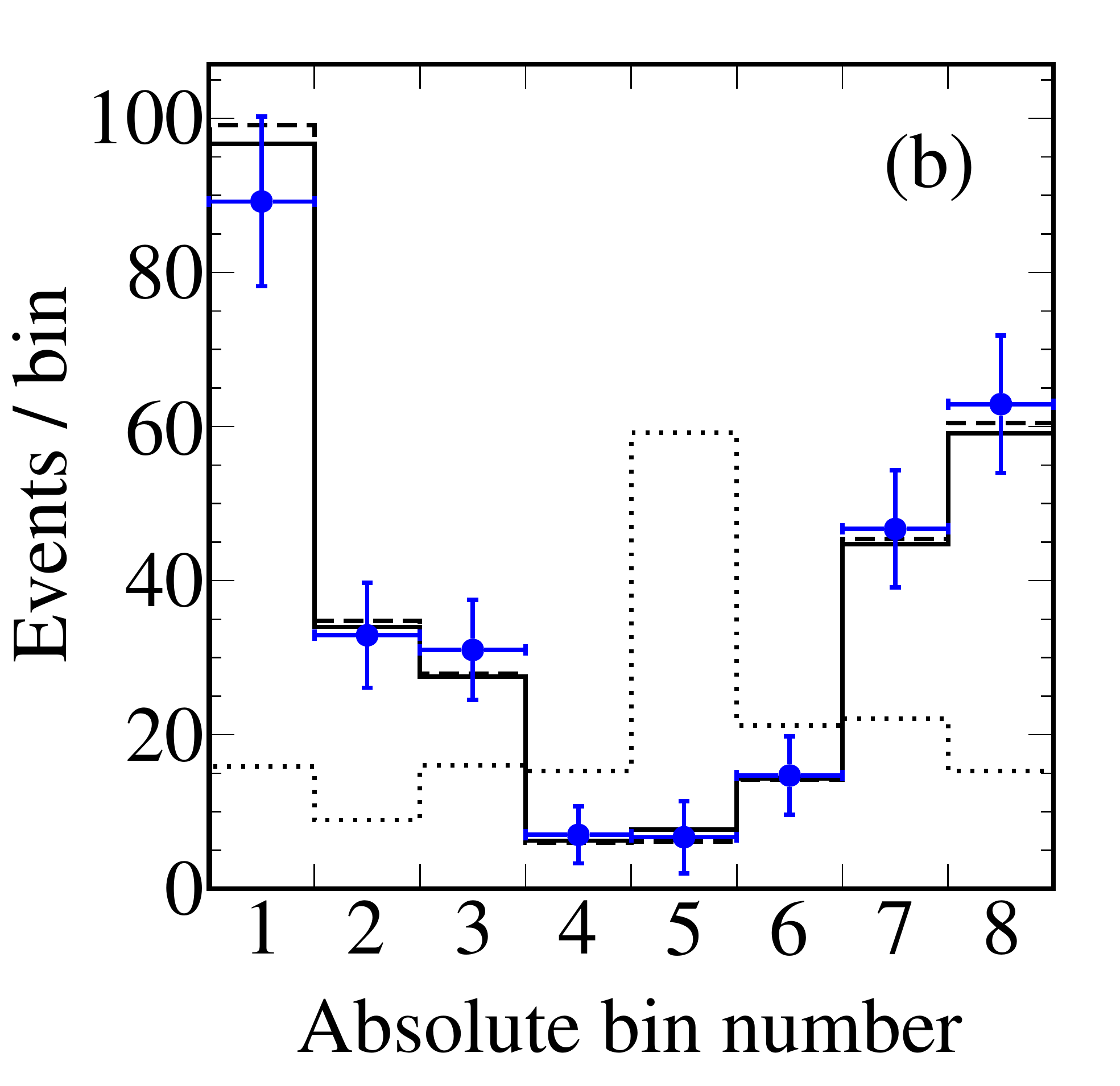} &
\includegraphics[width=0.32\columnwidth]{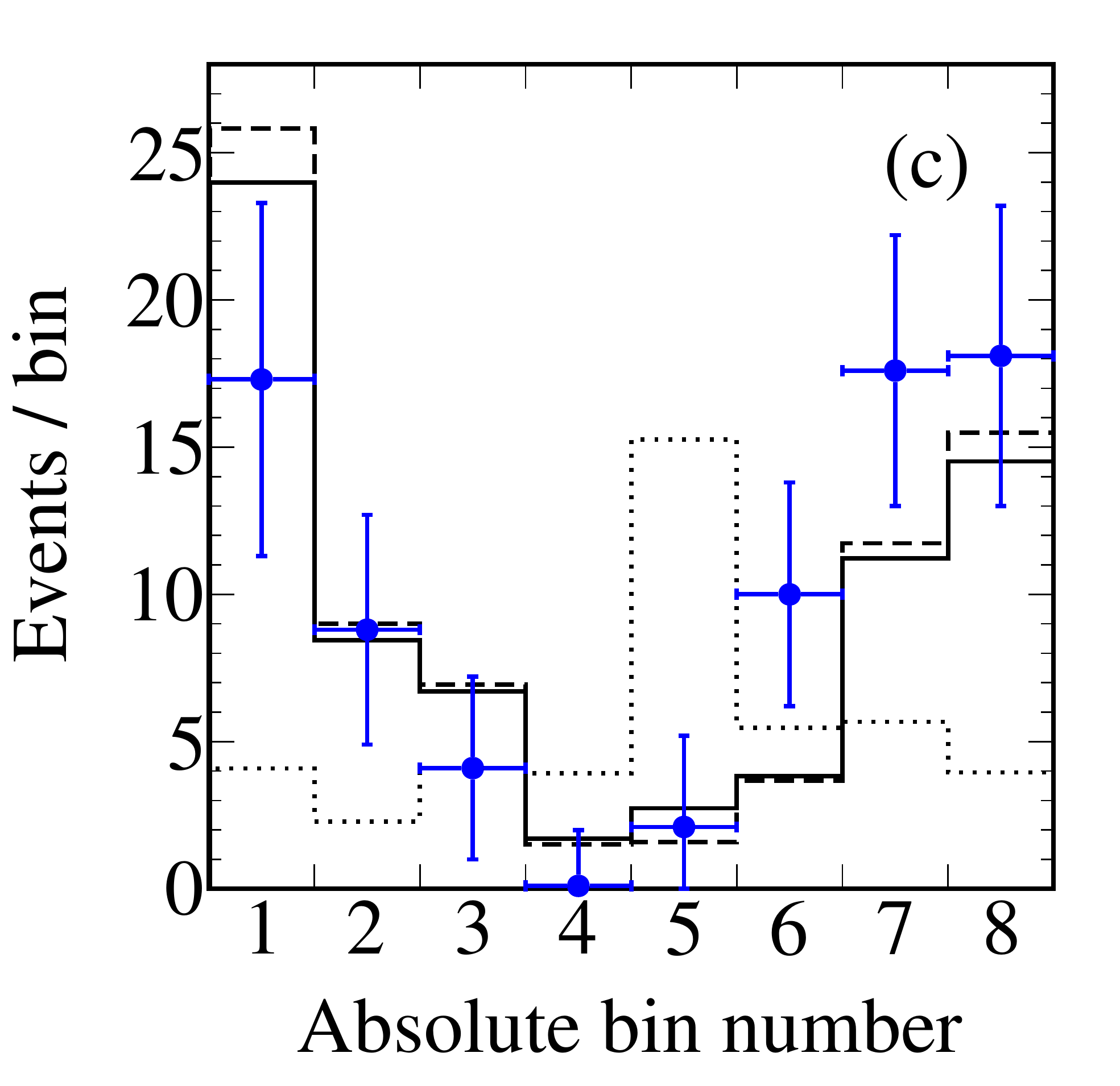} 
\end{tabular}
\caption{Data (points) and fit results (solid line) in absolute bin numbers for \Klpipi tags vs. (a) $D \to \pipipipi$, (b) $D \to 
\pipipiz$ and (c) $D \to \KKpiz$.  Also shown in each case is the expectation if $\Fp = 0$ (dotted line) or $\Fp = 1$ (dashed line). }  
\label{fig:fitklpipi}
\end{center}
\end{figure}

 A  log-likelihood  fit is performed to the efficiency-corrected signal yields of each sample, assuming the expected distributions given by Eqs.~\ref{eq:kspipi} and \ref{eq:klpipi}.  The fit parameters are the \CP-even fraction and the overall normalisation.  The values of $K_i$, $K_i^\prime$, $c_i$ and $c_i^\prime$ are also fitted, but with their measurement uncertainties and correlations imposed with Gaussian constraints.  Separate fits are performed for the $D \to K^0_{\rm S} \pi^+\pi^-$ tags, the $D \to K^0_{\rm L} \pi^+\pi^-$ tags, and for both samples combined.  Fits to large ensembles of simulated experiments demonstrate that the returned uncertainties are reliable and that there is no significant bias in the procedure. All data fits are found to be of good quality.  The results are plotted in Fig.~\ref{fig:fitkspipi} for the $D \to K^0_{\rm S} \pi^+\pi^-$ tags and in  Fig.~\ref{fig:fitklpipi} for the $D \to K^0_{\rm L} \pi^+\pi^-$ tags.  The numerical results for the \CP-even fraction are given in Table~\ref{tab:fourpik0pipi} for $D \to \pipipipi$ and in Table~\ref{tab:hhpi0k0pipi} for $D  \to \pi^+\pi^-\pi^0$ and $D \to K^+K^-\pi^0$.

\begin{table}[thb]
\begin{center}
\caption{The  \Ffourpi fit results for the $D \to K^0 \pi^+ \pi^-$ tags, where the first uncertainty is statistical and the second systematic.   The row $K^0_{\rm S,L} \pi^+\pi^-$ indicates the configuration where the \CP-even fraction is a common fit parameter shared  between the  $D \to K^0_{\rm S} \pi^+\pi^-$ and $D \to K^0_{\rm L} \pi^+\pi^-$ samples.} \label{tab:fourpik0pipi} \vspace*{0.1cm}
\begin{tabular}{ l r@{\ $\pm$\ }l@{\ $\pm$\ }l }  \hline \hline
 Tag                                        &  \multicolumn{3}{c}{\Ffourpi}       \\  \hline
$K^0_{\rm S} \pi^+\pi^-$    & 0.828 & 0.074 & 0.014   \\
$K^0_{\rm L} \pi^+\pi^-$    &  0.670 &  0.057  &  0.039  \\
$K^0_{\rm S,L} \pi^+\pi^-$    & 0.737 & 0.049  &  0.024 \\ \hline\hline
\end{tabular}
\end{center}
\end{table}

\begin{table}[thb]
\begin{center}
\caption{The  \Fppipipiz and \FpKKpiz fit results for the $D \to K^0 \pi^+ \pi^-$ tags, where the first uncertainty is statistical and the second systematic.   The row $K^0_{\rm S,L} \pi^+\pi^-$ indicates the configuration where the \CP-even fraction  is a common fit parameter shared  between the  $D \to K^0_{\rm S} \pi^+\pi^-$ and $D \to K^0_{\rm L} \pi^+\pi^-$ samples. }
\label{tab:hhpi0k0pipi}\vspace*{0.1cm}
\begin{tabular}{l r@{\ $\pm$\ }l@{\ $\pm$\ }l  r@{\ $\pm$\ }l@{\ $\pm$\ }l }  \hline \hline
 Tag                                     &  \multicolumn{3}{c}{\Fppipipiz}   &   \multicolumn{3}{c}{\FpKKpiz}    \\  \hline
$K^0_{\rm S} \pi^+\pi^-$   & 1.034 & 0.054  & 0.023 & 0.573 & 0.152 & 0.046  \\
$K^0_{\rm L} \pi^+\pi^-$   & 0.971 & 0.075  & 0.033 & 0.916 & 0.181 & 0.066   \\
$K^0_{\rm S,L} \pi^+\pi^-$   & 1.014 & 0.045 & 0.022 & 0.734 & 0.106 & 0.054  \\ \hline\hline
\end{tabular}
\end{center}
\end{table}

The dominant systematic uncertainty is associated with the distribution of the continuum and combinatoric backgrounds.  This is assessed by repeating the measurement with these contributions  
switched to those found in the sidebands of the signal distributions.  Shifts of $0.020$, $0.019$ and $0.037$ are observed for \Ffourpi, \Fppipipiz and \FpKKpiz, respectively. The uncertainty associated with the measurement errors on $K_i$, $K_i^\prime$, $c_i$ and $c_i^\prime$ is estimated by re-running the fit with these quantities set as fixed parameters and subtracting in quadrature the new fit uncertainty from that obtained with the original procedure.  This component  is found to be $0.013$ for $D \to \pipipipi$, $0.010$ for $D \to \pi^+\pi^-\pi^0$ and $0.025$ for $D \to K^+K^-\pi^0$, and is accounted as a systematic uncertainty in the final results.   An uncertainty is evaluated to account for non-uniformities in acceptance across phase space.  For $D \to \pipipipi$ this contribution is calculated with the same procedure as in Sect.~\ref{sec:cptags}, and found to be $0.002$ for the joint $K^0_{\rm S,L} \pi^+\pi^-$ fit.  For $D \to \pi^+\pi^-\pi^0$ and $D \to K^+K^-\pi^0$ the acceptance uncertainties are taken to be 0.001 and $0.010$, respectively, as determined in Ref.~\cite{MINAKSHI}. Other sources of bias are evaluated to be small.   The total systematic uncertainties are given in Tables~\ref{tab:fourpik0pipi} and~\ref{tab:hhpi0k0pipi} and are in all cases  significantly smaller than the corresponding statistical uncertainties.

\section{Other tags}
\label{sec:othertags}

The double-tagged yield of $\pipipipi$ vs. $\pi^+\pi^-\pi^0$ can be used to determine \Ffourpi, benefiting from the well-measured value  of \Fppipipiz. The ratio of double-tag  and $D \to \pi^+\pi^-\pi^0$ single-tag yields is defined as $N^{\pi\pi\pi^0}\equiv M(4\pi|\pi\pi\pi^0)/S({\pi\pi\pi^0})$, where a very small correction is applied to the measured single-tag yield to account for mixing effects. Following from Eq.~\ref{eq:Ndoubletag}, the ratio $N^{\pi\pi\pi^0}/N^{+}$ removes dependence on the signal branching fraction and reconstruction efficiency and is given by   
\begin{equation}
\frac{N^{\pi\pi\pi^0}}{N^+} = \frac{\left[ 1-\Big(2\Ffourpi -1\Big)\left(2\Fppipipiz-1\right)\right ]}{2\Ffourpi},  \label{eq:other1}
\end{equation}
which can be rearranged to yield
\begin{equation}
\label{F4pivshhpi0}
\Ffourpi= \frac{N^+\Fppipipiz}{N^{\pi\pi\pi^0} - N^+ +  2N^+\Fppipipiz}.
\end{equation}
The choice of $N^+$ in the denominator of Eq.~\ref{eq:other1} is preferred to $N^-$ as it is measured with better relative precision.

Taking as input the yields given in Sect.~\ref{sec:eventsel}, the value of $N^+$ reported in Sect.~\ref{sec:cptags}  and  the final result for \Fppipipiz  presented in Sect.~\ref{sec:combination} implies $\Ffourpi = 0.695 \pm 0.050 \pm 0.021$, where the uncertainties are statistical and systematic, respectively.   The main contributions to the systematic uncertainty arise from: the measurement of the $D \to \pi\pi\pi^0$ single-tag yield and small violations of the efficiency-factorisation ansatz assumed in Eq.~\ref{eq:other1}; the understanding of the peaking background component in the sample; and the possible effects of non-uniform acceptance.

The self-tagged yield of $\pipipipi$ vs. $\pipipipi$  also carries information on the value of \Ffourpi. This sample is however only used for a consistency check, as there are large backgrounds from both the continuum and from misidentification of $D \to K_{\rm S}^0\pi\pi$ decays that are a potential source of significant systematic bias. Furthermore, the predicted yield and measurement uncertainty means that the result from analysis of these double tags would have low weight in the combined measurement of \Ffourpi. Using Eq.~\ref{eq:Ndoubletag} the number of observed self-tagged events is given by
\begin{equation}
M(4\pi | 4\pi ) = 4\mathcal{R}\Ffourpi\Big(1-\Ffourpi\Big),
\end{equation}
where $\mathcal{R} = \NDDbar \mathcal{B}(4\pi)^2\varepsilon(4\pi|4\pi)$. The predicted double-tagged yield using the value of \Ffourpi obtained from the \CP tags is 17$\pm$2, which is consistent with the measured yield reported in Table~\ref{tab:doubletags}.

\section{Combination of results}
\label{sec:combination}

The results for \Fppipipipi from the \CP tags, the $K^0_{\rm S, L}\pi^+\pi^-$  tags and the \pipipiz tag are summarised in Table~\ref{tab:fourpisum}.  They are compatible and are therefore combined, taking account of correlated uncertainties.   Correlations arise from the non-flat Dalitz plot acceptance between all three measurements and the use of $N^+$ as an input to both the \CP tags and \pipipiz tag measurements.  There is a further small correlation between the results obtained with the \CP and \pipipiz tags, associated  with the uncertainty on the value of the mixing parameter $y_D$.  The final result is $\Fppipipipi = 0.737 \pm 0.028$.

\begin{table}[thb]
\begin{center}
\caption{Results for \Ffourpi for each tag category, and combined. When two uncertainties are shown, the first is statistical and the second systematic. For the combined result the total uncertainty is given. } \label{tab:fourpisum} \vspace*{0.1cm}
\begin{tabular}{ l r@{\ $\pm$\ }l@{\ $\pm$\ }l } \hline \hline
Tag & \multicolumn{3}{c}{\Ffourpi} \\ \hline
\CP eigenstates & 0.754 & 0.031 & 0.021 \\
$K^0_{\rm S,L} \pi^+\pi^-$ &0.737&0.049 &0.024 \\
$\pi^+\pi^-\pi^0$  & 0.695 &0.050 &0.021 \\ \hline
Combined & \multicolumn{3}{c}{0.737 $\pm$ 0.028} \\ \hline\hline
\end{tabular}
\end{center}
\end{table}

Table~\ref{tab:hhpi0sum} summarises the results on \Fppipipiz and \FpKKpiz obtained with $K^0_{\rm S, L}\pi^+\pi^-$ tags, together with those determined from \CP tags.
The $K^0_{\rm S, L}\pi^+\pi^-$ measurements confirm the results of the earlier analysis.  A combination is performed, accounting for the sole source of correlated uncertainties, which is that arising from the non-uniform acceptance over the Dalitz plots.   Results of $\Fppipipiz = 0.973 \pm 0.017$ and $\FpKKpiz = 0.732 \pm 0.055$ are obtained.  The $K^0_{\rm S, L}\pi^+\pi^-$  tags improve the relative precision on \Fppipipiz by 6\% and on \FpKKpiz by 10\%.

\begin{table}[thb]
\begin{center}
\caption{Results for \Fppipipiz and \FpKKpiz for each tag category, and combined. The \CP-eigenstate tag results are from Ref.~\cite{MINAKSHI}. When two uncertainties are shown, the first is statistical and the second systematic. For the combined result the total uncertainty is given. }\label{tab:hhpi0sum} \vspace*{0.1cm}
\begin{tabular}{l r@{\ $\pm$\ }l@{\ $\pm$\ }l r@{\ $\pm$\ }l@{\ $\pm$\ }l} \hline \hline
Tag & \multicolumn{3}{c}{\Fppipipiz} & \multicolumn{3}{c}{\FpKKpiz} \\ \hline
\CP eigenstates & 0.968 & 0.017 & 0.006 & 0.731 & 0.058 & 0.021 \\
$K^0_{\rm S,L} \pi^+\pi^-$ & 1.014 &0.045 & 0.022&0.734 &0.106 & 0.054 \\ \hline
Combined & \multicolumn{3}{c}{0.973 $\pm$ 0.017} & \multicolumn{3}{c}{0.732 $\pm$ 0.055} \\ \hline\hline
\end{tabular}
\end{center}
\end{table}

\section{Conclusions}
\label{sec:conc}

A first measurement has been made of  the \CP-even fraction of the decay $D \to \pipipipi$, exploiting quantum-correlated double-tags involving \CP-eigenstates, a binned Dalitz-plot analysis of the modes $D \to K^0_{\rm S,L}\pi^+\pi^-$, and $D \to \pipipiz$ decays.  The result, $\Fppipipipi = 0.737 \pm 0.028$, when considered alongside the relatively high branching fraction, indicates that this channel is a valuable addition to the suite of $D$ decays that can be harnessed for the measurement of the unitarity-triangle angle $\gamma$ through the process $B^\mp \to DK^\pm$.
The decays $D \to K^0_{\rm S,L}\pi^+\pi^-$ have also been employed as a tag to measure the \CP contents of the modes $D \to \pipipiz$ and $D \to \KKpiz$. The results confirm the conclusion of a previous analysis~\cite{MINAKSHI}, based on \CP-eigenstate tags, and also suggested by earlier amplitude-model studies~\cite{CLEOPIPIPI0,BaBarPIPIPI0,BRIAN}, that the \CP-even content of the $\pipipiz$ final state is very high, and therefore this decay too is a powerful mode for the measurement of $\gamma$. Combining the two sets of measurements yields $\Fppipipiz = 0.973 \pm 0.017$ and $\FpKKpiz = 0.732 \pm 0.055$. 
Now that their \CP-even fractions have been measured, all three decay modes may also be used for studies of indirect \CP violation and mixing in the \Dz\Dzb system~\cite{SNEHACHRISGUY}.

\section*{Acknowledgments}

This analysis was performed using CLEO-c data, and as members of the former CLEO collaboration we thank it for this privilege.
We are grateful for support from the UK Science and Technology Facilities Council, the UK-India  Education and Research Initiative,
and the European Research Council under FP7.

\end{document}